\pdfoutput=1
\documentclass[a4paper,12pt]{article}
\usepackage[utf8]{inputenc}
\usepackage[top=50pt,bottom=60pt,left=68pt,right=66pt]{geometry}
\usepackage{amsmath}
\usepackage{booktabs}
\usepackage{amssymb}
\usepackage[title]{appendix}
\usepackage{mathrsfs}
\usepackage{float}
\usepackage{xcolor}
\usepackage{multirow}
\usepackage{graphicx,caption,subcaption}
\usepackage{comment}
\usepackage[space]{grffile}
\newcommand{\overbar}[1]{\mkern1.5mu\overline{\mkern-1.5mu#1\mkern-1.5mu}\mkern 1.5mu}

\def\un#1{\relax\ifmmode\@@underline#1\else
        $\@@underline{\hbox{#1}}$\relax\fi}

\let\du=\du                     

\def\s{\sigma}

\def\D{\Delta}

\def\O{\Omega}

\def\co{{\cal O}}

\def\car{{\cal R}}

\def\bo{{\raise-.3ex\hbox{\large$\Box$}}}               
\def\TH{{\raise.2ex\hbox{$\displaystyle \bigodot$}\mskip-4.7mu \llap H \;}}
\def\face{{\raise.2ex\hbox{$\displaystyle \bigodot$}\mskip-2.2mu \llap {$\ddot
        \smile$}}}                                      

\def\leftrightarrowfill{$\mathsurround=0pt \mathord\leftarrow \mkern-6mu
        \cleaders\hbox{$\mkern-2mu \mathord- \mkern-2mu$}\hfill
        \mkern-6mu \mathord\rightarrow$}
\def\dvec#1{\vbox{\ialign{##\crcr
        \leftrightarrowfill\crcr\noalign{\kern-1pt\nointerlineskip}
        $\hfil\displaystyle{#1}\hfil$\crcr}}}           

\def\frac#1#2{{\textstyle{#1\over\vphantom2\smash{\raise.20ex
        \hbox{$\scriptstyle{#2}$}}}}}                   
\def\sfrac#1#2{{\vphantom1\smash{\lower.5ex\hbox{\small$#1$}}\over
        \vphantom1\smash{\raise.4ex\hbox{\small$#2$}}}} 
\def\bfrac#1#2{{\vphantom1\smash{\lower.5ex\hbox{$#1$}}\over
        \vphantom1\smash{\raise.3ex\hbox{$#2$}}}}       
\def\afrac#1#2{{\vphantom1\smash{\lower.5ex\hbox{$#1$}}\over#2}}    

\def\[{\lfloor{\hskip 0.35pt}\!\!\!\lceil}
\def\]{\rfloor{\hskip 0.35pt}\!\!\!\rceil}

\def\du#1#2{_{#1}{}^{#2}}

\def\un{\underline}
\def\fracmm#1#2{{{#1}\over{#2}}}

\def\low#1{{\raise -3pt\hbox{${\hskip 0.75pt}\!_{#1}$}}}

\newskip\humongous \humongous=0pt plus 1000pt minus 1000pt

\newif\ifdtup

\def\({\left(}
\def\){\right)}

\def\beq{\begin{equation}}
\def\eeq{\end{equation}}
\def\bea{\begin{eqnarray}}
\def\eea{\end{eqnarray}}

\newcommand{\be}{\begin{equation}}
\newcommand{\ee}{\end{equation}}
\newcommand{\nbe}{\begin{equation*}}
\newcommand{\nee}{\end{equation*}}

\begin{document}
\renewcommand{\arraystretch}{1.3}

\thispagestyle{empty}


\noindent {\hbox to\hsize{
\vbox{\noindent July 2022 \hfill IPMU22-0033 }}
\noindent  \hfill }

\noindent
\vskip2.0cm
\begin{center}

{\Large\bf Inflation, SUSY breaking, and primordial black holes in modified supergravity 
coupled to chiral matter}

\vglue.3in

Yermek Aldabergenov~${}^{a,b}$, Andrea Addazi~${}^{c,d}$, and Sergei V. Ketov~${}^{e,f,g}$
\vglue.3in

${}^a$~Department of Physics, Faculty of Science, Chulalongkorn University\\
Thanon Phayathai, Pathumwan, Bangkok 10330, Thailand\\
${}^b$~Department of Theoretical and Nuclear Physics, Al-Farabi Kazakh National University, 71 Al-Farabi Avenue, Almaty 050040, Kazakhstan\\
${}^c$~Center for Theoretical Physics, College of Physics, Science and Technology \\
Sichuan University,  610065 Chengdu,  China\\
${}^d$~Laboratori Nazionali di Frascati (INFN), Via Enrico Fermi 54 \\
 I-00044 Frascati (Rome), Italy \\
${}^e$~Department of Physics, Tokyo Metropolitan University\\
1-1 Minami-ohsawa, Hachioji-shi, Tokyo 192-0397, Japan \\
${}^f$~Research School of High-Energy Physics, Tomsk Polytechnic University\\
2a Lenin Avenue, Tomsk 634028, Russian Federation\\
${}^g$~Kavli Institute for the Physics and Mathematics of the Universe (WPI)
\\The University of Tokyo Institutes for Advanced Study, Kashiwa 277-8583, Japan\\
\vglue.1in

 yermek.a@chula.ac.th, addazi@scu.edu.cn, ketov@tmu.ac.jp
\end{center}

\vglue.3in

\begin{center}
{\Large\bf Abstract}
\end{center}

We propose a novel model of the modified (Starobinsky-like) old-minimal-type supergravity coupled to a chiral matter superfield, that can {\it simultaneously} describe multi-field inflation, primordial black hole (PBH) formation, dark matter (DM), and spontaneous supersymmetry (SUSY) breaking after inflation in a Minkowski vacuum.  The PBH masses in our supergravity model of double slow-roll inflation, with a short phase of "ultra-slow-roll" between two slow-roll phases, are close to $10^{18}$ g. We find that a significant PBH fraction in the allowed mass window can be supplemented by spontaneous SUSY breaking in the vacuum with the gravitino mass close to the scalaron (inflaton) mass M of the order $10^{13}$ GeV. Our supergravity model favors the {\it composite} nature of  DM as a mixture of PBH and heavy gravitinos as the lightest SUSY particles. The composite DM significantly relaxes fine-tuning needed for the whole PBH-DM. The
PBH-DM fraction is derived, and the second-order gravitational wave background induced by the enhanced scalar perturbations is calculated. Those gravitational waves may be accessible by the future space-based gravitational interferometers.

\newpage

\section{Introduction}

The origin and evolution of black holes is one of the actual issues in modern physics. The standard (astrophysical) origin of black holes due to gravitational collapse of stars leads to the lower (Chandrasekhar) bound on the black hole mass (above three Solar masses).  However, then it is difficult to explain the origin of supermassive black holes  at present times and the origin of black holes already in the early half of the first Gyear of our Universe,
whose existence is supported by current astronomical observations.

The alternative idea about the primordial origin of black holes was proposed by Zeldovich and Novikov \cite{Novikov:1967tw}, and then by Hawking \cite{Hawking:1971ei}. The idea of primordial black holes (PBH) can be naturally coupled to the inflationary paradigm because cosmological inflation can be considered as the most powerful natural accelerator that can produce very large primordial density perturbations as the seeds of PBH of any mass, also well below the Solar mass \cite{Dolgov:1992pu}.  More recently, the PBH proposal was supported by the NANOGrav observations \cite{Arzoumanian:2020vkk}. 

There is the related proposal that PBH can form 
the (non-particle) dark matter (DM) \cite{Barrow:1992hq,Ivanov:1994pa}, which attracted a lot of attention in the recent literature (see, for instance, Refs.~\cite{Cai:2018dig,Bartolo:2018evs,DeLuca:2020agl} and the references therein), provided that the PBH masses are above the Hawking (black hole) radiation limit of $10^{15}$ g, so that those PBH can survive in the present Universe. The two ideas (PBH and PBH-DM) are intrinsically independent because PBH may represent merely part of DM or none of it. Even when PBH never formed, their theoretical studies are relevant because they constrain physics of the very early Universe \cite{Carr:2003bj}. The observational constraints on PBH and PBH-DM are reviewed in Refs.~\cite{Sasaki:2018dmp,Carr:2020gox,Carr:2020xqk}. In particular, the whole PBH-DM is only possible with the PBH masses between $10^{-15}$ and $10^{-12}$ times the Solar mass.

As regards cosmological inflation and  PBH, the key theoretical question is about the PBH formation mechanism.  The expected scale of inflation (about $10^{13}$ GeV) implies that the PBH formation belongs to super-high energy physics well beyond the electroweak scale, which requires the model building based on new physics at those scales in agreement with current observations of the cosmic microwave background (CMB) radiation. We assume that the new physics is given by supersymmetry (SUSY) that is the standard proposal well beyond the Standard Model (SM) of elementary particles. In the context of inflation and the very early Universe, it means the $N=1$ supergravity framework
in four spacetime dimensions. We employ the PBH formation mechanism based on gravitational instabilities induced by scalar fields \cite{M1}, which are quite natural in the supergravity framework \cite{Ketov:2021fww}.

In our earlier papers \cite{Aldabergenov:2020bpt,Aldabergenov:2020yok,Ishikawa:2021xya}, the modified (Starobinsky-like) supergravity was applied to describe the PBH formation after Starobinsky's inflation for the first time.~\footnote{See also Refs.~\cite{Dalianis:2018frf,Nanopoulos:2020nnh,Mahbub:2021qeo,Stamou:2021qdk} about PBH formation in more general frameworks of no-scale supergravity and superconformal $\alpha$-attractors.}  We define the modified supergravity as the non-perturbative supergravity extension of the $(R+R^2)$ gravity with linearly realized local SUSY  \cite{Ketov:2010qz,Ketov:2012jt,Ketov:2013dfa,Addazi:2017rkc}. As was demonstrated in Ref.~\cite{Aldabergenov:2020bpt},  the Starobinsky supergravity can lead to the effective two-(scalar)-field attractor-type double inflation, whose first stage is driven by Starobinsky's inflaton (scalaron) and whose second stage is driven by another scalar that belongs to the full (old-minimal) supergravity multiplet. The large primordial perturbations ($10^6\div 10^7$ times higher than those of CMB) can be generated by the "isocurvature pumping" mechanism between the two stages  of inflation, resulting in the so-called "ultra-slow-roll" regime \cite{Germani:2017bcs}. The isocurvature pumpling is robust, and it is possible only in multi-field inflation  \cite{GarciaBellido:1996qt,Kawasaki:2015ppx,Braglia:2020eai,Fumagalli:2020adf,Gundhi:2020kzm}. It is the ultra-slow-roll regime with the negative isocurvature mass squared that leads to an enhancement (a large peak) in the power spectrum of scalar perturbations, which, in turn, yields an efficient formation of PBH.  The resulting PBH masses were estimated in the range between $10^{16}$ g and $10^{20}$ g \cite{Aldabergenov:2020bpt} with a possibility of the whole PBH-DM. It was also found that the gravitational waves (GW) induced by the PBH formation in the supergravity framework may be detected by the future space-based gravitational interferometers such as LISA and DECIGO \cite{Aldabergenov:2020yok}.
 
The Starobinsky model \cite{Starobinsky:1980te} of inflation (see e.g., Ref.~\cite{Ketov:2019toi} for a recent review) well fits current CMB observations, relies only on gravitational interactions, while its only (inflaton mass) parameter $M$ is fixed by the CMB amplitude. Starobinsky's inflaton (scalaron) is the physical degree of freedom of the higher-derivative gravity. It can be identified with the Nambu-Goldstone boson related to spontaneous breaking of the scale invariance \cite{Ketov:2012jt}. The current observational constraints on cosmological  inflation can be found in Refs.~\cite{Planck:2018jri,BICEP:2021xfz,Tristram:2021tvh}.

The minimal supergravity extension of modified $F(R)$ gravity models with the linearly (manifestly) realized SUSY selects the Starobinsky inflation model \cite{Starobinsky:1980te} based on the $(R+R^2)$ gravity, relates inflation to PBH production and DM genesis, and has more physical scalars (in addition to scalaron) because the "auxiliary" fields of the (old-minimal) supergravity multiplet become physical or propagating \cite{Ketov:2012jt,Ketov:2013dfa,Addazi:2017rkc} thus leading to multi-field inflation. The models of modified supergravity proposed and studied in 
Refs.~\cite{Aldabergenov:2020bpt,Aldabergenov:2020yok,Ishikawa:2021xya} can always be rewritten to the scalar-tensor form without higher derivatives, like $F(R)$ gravity models. However, those modified supergravity models also lead to a limited tension (within $3\s$) with the observed value \cite{Planck:2018jri} of the CMB tilt $n_s$ of scalar perturbations and have only Minkowski vacua after inflation, where SUSY is restored. 

Therefore, it is desirable to find possible generalizations or extensions of the modified supergravity models \cite{Aldabergenov:2020bpt,Aldabergenov:2020yok,Ishikawa:2021xya} of inflation and PBH production. The common feature of all those models is their minimal field content that is entirely based on a single supergravity multiplet, with all its fields being related via SUSY transformations. On the one side, these models are highly restrictive and may be easily falsified, though being powerful enough to explain inflation, PBH production and PBH-DM without matter fields different from supergravity, by using only the "pure" supergravity fields and their interactions. On the other side, the restrictive features of those models may be relaxed by adding SUSY matter to the modified supergravity. It opens a new avenue in the model building of inflation, PBH and DM, all based on supergravity. Of course, there are many choices of adding matter, like in the standard supergravity theory. In this paper we merely consider adding a single chiral matter multiplet (or a chiral superfield) because our primary interest is in the scalar sector of the new supergravity theories. We use the curved superspace formalism  for describing supergravity (in the old-minimal version) with the standard notation \cite{Wess:1992cp}.  We also set the reduced Planck mass $M_{\rm Pl}=1$ for simplicity, unless stated otherwise.

Our paper is organized as follows. Our setup for the modified Starobinsky-like supergravity coupled to  chiral matter, and the manifestly supersymmetric Lagrangians of our new models are given in Sec.~2, where we also calculate their scalar-tensor part, i.e. the {\it derived } Lagrangians of scalars coupled to gravity in the Einstein frame. In Sec.~3 we introduce a simplified two-field model (ignoring other scalars) from the modified supergravity as our startup, in order to get insights into inflationary dynamics. Our full new model is introduced and investigated in Sec.~4 where the contribution of all relevant (for inflation) scalars is taken into account. Spontaneous SUSY breaking in our model after inflation in a Minkowski vacuum is derived in Sec.~5. The PBH fraction in DM for the full model is computed in Sec.~6. The gravitational waves (GW) induced by scalar perturbations related to the PBH formation are briefly studied in Sec.~7. Our conclusion is Sec.~8. 

\section{Modified supergravity coupled to chiral matter}\label{sec_setup}

The (manifestly supersymmetric) curved superspace Lagrangian of a chiral matter superfield $\Phi$ minimally coupled to the modified supergravity is given by 
\begin{equation}
    {\cal L}=\int d^2\Theta 2{\cal E}\left[-\frac{1}{8}(\overbar{\cal D}^2-8{\cal R})(N+J)+{\cal F}+\Omega\right]+{\rm h.c.} \label{L_master_1}
\end{equation}
It is parametrized by four arbitrary potentials: two non-holomorphic ones $N=N({\cal R},\overbar{\cal R})$ and  $J=J(\Phi,\overbar\Phi)$, and two holomorphic ones ${\cal F}={\cal F}({\cal R})$ and $\Omega=\Omega(\Phi)$, as functions of the chiral scalar curvature superfield ${\cal R}$ of supergravity and the chiral superfield $\Phi$ of matter.

The standard (Poincar\'e or Einstein) supergravity is recovered by setting $N=0$ and ${\cal F}=-3{\cal R}$ (alternatively, one can set ${\cal F}=0$ and $N=-3$) with $J =0$ and $\O=0$.
  Given generic potentials $N$ and ${\cal F}$, the Lagrangian in terms of the field components has only the first and second powers of the spacetime scalar curvature $R$ that appears in the
coefficient of the ${\cal R}$-superfield at $\Theta^2$. That is why we call it the Starobinsky-like Lagrangian of modified supergravity. When $\Phi$ enters $J$ and $\Omega$ without mixing with $\cal R$, we call it the minimal coupling of the modified supergravity to chiral matter, as in Eq.~(\ref{L_master_1}).

Eliminating the auxiliary $F$-field of $\Phi$ in terms of the leading scalar field components $\phi\equiv \Phi|$ and $X\equiv{\cal R}|$ yields
\begin{equation}
	F=-J^{-1}_{,\phi\bar\phi}(2\overbar XJ_{,\bar\phi}+\overbar\Omega_{,\bar\phi})~.\label{F_term}
\end{equation}
where the subscripts with commas denote the derivatives with respect to the given (field) arguments. 

A tedious calculation with vanishing fermionic fields yields the bosonic part of the Lagrangian as follows:
\begin{align}
\begin{split}
	e^{-1}{\cal L}_{\rm bos.}= &-\tfrac{1}{12}\big({\cal F}_{,X}+\overbar{\cal F}_{,\bar X}+2N_{,X}X+2N_{,\bar X}\overbar X-8N_{,X\bar X}X\overbar X+2N+2J-\tfrac{1}{9}N_{,X\bar X}b_mb^m\big)R\\
	&+\tfrac{1}{144}N_{,X\bar X}R^2-N_{,X\bar X}\partial X\partial\overbar X-J_{,\phi\bar\phi}\partial\phi\partial\bar\phi-\tfrac{i}{3}b_m(N_{,X}\partial^mX+J_{,\phi}\partial^m\phi-{\rm c.c.})\\
	&+\tfrac{i}{6}\big({\cal F}_{,X}-\overbar{\cal F}_{,\bar X}+2N_{,X}X-2N_{,\bar X}\overbar X
	-\tfrac{i}{6}N_{,X\bar X}D_mb^m\big)D_mb^m\\
	&-\tfrac{1}{2}\big({\cal F}_{,X}+\overbar{\cal F}_{,\bar X}+2N_{,X}X+2N_{,\bar X}\overbar X-4N_{,X\bar X}X\overbar X+2N+2J\\
	&\qquad-\tfrac{1}{18}N_{,X\bar X}b_mb^m\big)(8X\overbar X+\tfrac{1}{9}b_mb^m)+6X(\overbar{\cal F}+\overbar\Omega)+6\overbar X({\cal F}+\Omega)\\
	&+12X\overbar X(N+J)-J^{-1}_{,\phi\bar\phi}|2XJ_{,\phi}+\Omega_{,\phi}|^2~.\label{L_master_comp}
\end{split}
\end{align}
When ignoring also the vector field, $b_m=0$, the Lagrangian above can be rewritten to the form 
\begin{equation} \label{JordanL}
	e^{-1}{\cal L}_{\rm bos.}=\fracmm{A}{2}R+\fracmm{B}{12M^2}R^2-\fracmm{12B}{M^2}\partial X\partial\overbar X-J_{,\phi\bar\phi}\partial\phi\partial\bar\phi-U~,
\end{equation}
with the specific scalar potential $U$, where we have also defined
\begin{equation} \label{Bdef}
B\equiv \tfrac{1}{12}M^2N_{,X\bar X}~.
\end{equation}
 The mass parameter $M$ has been introduced here for later convenience. Other quantities are given by
\begin{align}
A &\equiv -\tfrac{1}{6}\big({\cal F}_{,X}+\overbar{\cal F}_{,\bar X}+2N_{,X}X+2N_{,\bar X}\overbar X-8N_{,X\bar X}X\overbar X+2N+2J\big)~,\\
U &\equiv 4X\overbar X\big({\cal F}_{,X}+\overbar{\cal F}_{,\bar X}+2N_{,X}X+2N_{,\bar X}\overbar X-4N_{,X\bar X}X\overbar X-N-J\big)\nonumber\\
&\hspace{2cm}-6X(\overbar{\cal F}+\overbar\Omega)-6\overbar X({\cal F}+\Omega)+J^{-1}_{,\phi\bar\phi}|2XJ_{,\phi}+\Omega_{,\phi}|^2~.\label{U_def}
\end{align}

The Lagrangian (\ref{JordanL}) in the Jordan frame can be rewritten to the dual (scalar-tensor) Lagrangian with the scalaron field $\varphi$ in the Einstein frame \cite{Whitt:1984pd,Maeda:1988ab}. We find
\begin{align}
\begin{aligned}
	e^{-1}{\cal L}_{\rm bos.}=\frac{1}{2}R-\frac{1}{2}\partial\varphi\partial\varphi-e^{-\sqrt{\frac{2}{3}}\varphi}\Big(\fracmm{12B}{M^2}\partial X\partial\overbar X+J_{,\phi\bar\phi}\partial\phi\partial\bar\phi\Big)\\
	-\fracmm{3M^2}{4B}\Big(1-Ae^{-\sqrt{\frac{2}{3}}\varphi}\Big)^2-e^{-2\sqrt{\frac{2}{3}}\varphi}U~.\label{L_FN_min}
\end{aligned}
\end{align}

We fix the supergravity potentials in the minimalistic way as follows:
\begin{equation}
	{\cal F}=-3X~,\quad N=\fracmm{12}{M^2}X\overbar X-\fracmm{72}{M^4}\zeta(X\overbar X)^2~,\label{FN_choice}
\end{equation}
just needed for the proper embedding of the Starobinsky $(R+R^2)$ gravity model of inflation into the modified supergravity  \cite{Addazi:2017rkc}. In particular, the parameter $M$ is proportional to the scalaron mass $m_{\varphi}$ as $m^2_{\varphi}=M^2/\langle B\rangle$ after assuming that $\langle Ae^{-\sqrt{2/3}\varphi}\rangle=1$ and $\langle U\rangle=0$, where the angle brackets denote the vacuum expectation values (VEV). The extra (second) term in $N$ with the real parameter $\zeta>0$ is needed for stabilization of the inflationary trajectory and the vacuum \cite{Addazi:2017rkc}.

After the rescalings
\begin{equation}
    X\rightarrow MX/\sqrt{12}~,\quad \Omega\rightarrow M\Omega/\sqrt{3}~,
\end{equation}
the Lagrangian takes the form
\begin{align}
\begin{aligned}
	e^{-1}{\cal L}_{\rm bos.}=\frac{1}{2}R-\frac{1}{2}\partial\varphi\partial\varphi-e^{-\sqrt{\frac{2}{3}}\varphi}(B\partial X\partial\overbar X+J_{,\phi\bar\phi}\partial\phi\partial\bar\phi)\\
	-\fracmm{3M^2}{4B}\Big(1-Ae^{-\sqrt{\frac{2}{3}}\varphi}\Big)^2-e^{-2\sqrt{\frac{2}{3}}\varphi}U~.\label{L_FN_norm}
\end{aligned}
\end{align}
where the functions $A,B,U$ read
\begin{gather}
\begin{gathered}
A=1+\tfrac{1}{3}(X\overbar X-J)-\tfrac{11}{6}\zeta(X\overbar X)^2~,\qquad B=1-2\zeta X\overbar X~,\\
U=M^2\left[X\overbar X\left(1-\tfrac{1}{3}J\right)-\tfrac{1}{3}(X\overbar X)^2+\tfrac{3}{2}\zeta(X\overbar X)^3-X\overbar\Omega-\overbar X\Omega+\tfrac{1}{3}J^{-1}_{,\phi\bar\phi}|XJ_\phi+\Omega_{,\phi}|^2\right]~.\label{U_def_FN_choice}
\end{gathered}
\end{gather}
It is worth emphasizing that Eqs.~(\ref{L_FN_norm}) and (\ref{U_def_FN_choice}) were {\it derived} from supergravity, and not postulated. The Lagrangian (\ref{L_FN_norm})  has the form of a non-linear sigma-model
(NLSM) \cite{Ketov:2000dy} in terms of two complex scalars $(X,\phi)$ and one real scalaron $\varphi$, having the full scalar potential $V$ given by the last two terms in  (\ref{L_FN_norm}), 
and minimally coupled to Einstein gravity (i.e., in the Einstein frame). The scalar potential $U$ is associated
with the Jordan frame, as is clear from Eq.~(\ref{JordanL}).

\section{Double inflation in a simplified model}\label{sec_demo}

In this paper, we are interested not only in a viable realization of inflation and a PBH production in supergravity, but also in spontaneous SUSY breaking after inflation at a high-energy scale comparable with the scale of inflation, in order to make contact with low-energy particle phenomenology. From the viewpoint of SUSY breaking, the chiral superfield we have introduced in the preceding Section can be considered as (part of) the hidden sector where SUSY breaking occurs.~\footnote{In modified supergravity, SUSY can also be broken by supergravity multiplet only \cite{Dalianis:2014aya}, but it makes it difficult to incorporate PBH production, so we do not consider this possibility.}

The simplest model of SUSY breaking in the Einstein supergravity with a single chiral matter superfield is known as the {\it Polonyi} model \cite{Polonyi:1977pj}. In the Polonyi model, supergravity is minimally coupled to a chiral matter superfield $\Phi$ with the canonical kinetic term and the superpotential given by a {\it linear} polynomial in $\Phi$. It leads to a Minkowski vacuum with spontaneously broken SUSY at any scale. In the phenomenological applications, the Polonyi superfield is usually affiliated with the hidden (heavy) matter sector that interacts with the observable matter (like the Standard Model) only gravitationally.  The SUSY breaking is supposed to be transferred to the observable sector at the electro-weak scale by gravity mediation. It is, therefore, natural to employ a Polonyi-like chiral superfield in our framework of Sec.~2 for inflation, PBH formation and SUSY breaking.

Surprisingly, we find that the standard Polonyi model in its simplest form does {\it not} work for SUSY breaking in the modified supergravity. Therefore, we have to modify the original Polonyi model  \cite{Polonyi:1977pj} by allowing more general superpotentials and/or K\"ahler potentials.  On the one hand, as can be seen from Eq.~\eqref{U_def} when $X=0$, a Minkowski vacuum requires $\Omega_{,\phi}=0$. This leads to $F=0$ from Eq.~\eqref{F_term}. Hence, in order to break SUSY in a  Minkowski vacuum, we necessarily need a non-vanishing vacuum expectation value (VEV) of $X$, $\langle X\rangle\neq 0$. On the other hand, if at the onset of inflation the fields $\phi$ and $X$ are stabilized around zero, as is the case in our models, there will be a non-trivial multi-field dynamics at smaller (than CMB) scales when the Jordan frame potential $U(X,\overbar X,\phi,\bar\phi)$ starts to control the dynamics (assuming it is initially suppressed), and the inflationary trajectory turns towards the non-vanishing VEV of $X$ and $\phi$. It makes the story truly complicated.
 
Therefore, before considering full dynamics of interacting fields $\varphi$, $X$, and $\phi$ in the next Section, in this Section we study the simplified two-field model where we set $X=0$ and consider only the real part of $\phi$. It does  not lead to SUSY breaking in the vacuum but provides some important insights into the multi-field inflation in our models.
 
Let us consider an extension of the Polonyi model  with the canonical kinetic term and the Wess-Zumino-type superpotential,
\begin{equation}
    J=\phi\bar\phi~,~~~\Omega=b\phi+\tfrac{c}{2}\phi^2+\tfrac{f}{3}\phi^3~,
\end{equation}
and the real parameters $\{b,c,f\}$. Having fixed $J$ and $\Omega$, we introduce the real scalar  $\rho\equiv\sqrt{2}\phi$, ignore the imaginary part of $\phi$ for simplicity, and set $X=0$ in 
Eq.~\eqref{L_FN_norm}. It yields the Lagrangian 
\begin{equation}
	e^{-1}{\cal L}=\tfrac{1}{2}\left(R-\partial\varphi\partial\varphi-e^{-\sqrt{\frac{2}{3}}\varphi}\partial\rho\partial\rho\right)-\tfrac{3}{4}M^2\left[1-\left(1-\tfrac{1}{6}\rho^2\right)e^{-\sqrt{\frac{2}{3}}\varphi}\right]^2-e^{-2\sqrt{\frac{2}{3}}\varphi}U~,\label{L_FN_toy}
\end{equation}
where the potential $U$ reads ({\it cf.} Refs.~\cite{Aldabergenov:2020bpt,Pi:2017gih})
\begin{equation}
    M^{-2}U=\tfrac{1}{3}\Omega_{,\phi}\overbar\Omega_{,\bar\phi}=\tfrac{1}{3}\left(b+\tfrac{c}{\sqrt{2}}\rho+\tfrac{f}{2}\rho^2\right)^2~.\label{U_toy}
\end{equation}

\begin{figure}
\centering
  \centering
  \includegraphics[width=.9\linewidth]{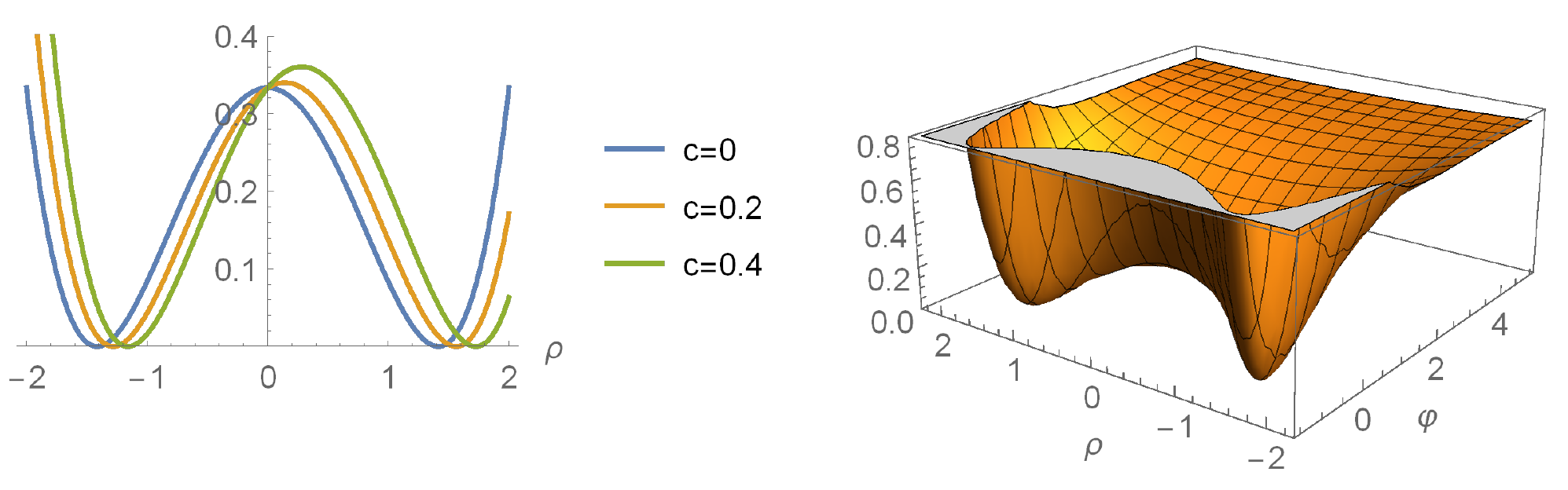}
\captionsetup{width=1\linewidth}
\caption{The single-field scalar potential $U/M^2$ in Eq.~\eqref{U_toy} for some values of the parameter $c$ (on the left), and the two-field scalar potential $V/M^2$ in Eq.~\eqref{L_FN_toy} for $c=0$ (on the right). The other parameters used are $b=-f=1$.}\label{Fig_toy_UV}
\end{figure}

When $c=0$ and $bf<0$, the potential $U$ is Higgs-like, and has two minima at $\rho=\pm\sqrt{-2b/f}$ and a local maximum at the origin. The coefficient  $c\neq 0$ deforms the potential as is shown on the left of Fig.~\ref{Fig_toy_UV}. Then the local maximum is shifted away from the origin, which can be used to our advantage, as is explained below. The minima in the presence of a non-vanishing $c$ are given by
\begin{equation}
    \langle\rho\rangle=-\fracmm{c}{\sqrt{2}f}\left(1\pm\sqrt{1-\fracmm{4bf}{c^2}}\right)~.
\end{equation}

Let us first describe what happens with the inflationary solution when $c=0$. When scalaron is large, $\varphi\gg 1$, the universe undergoes the regular Starobinsky-like inflation because the $U$ is suppressed against the Starobinsky-like potential given by the first term in the scalar potential \eqref{L_FN_toy}, when assuming the parameters $b,c,f$ to be not too large. On the other hand, the scalar $\rho$ is stabilized around zero with the effective mass
\begin{equation}
    m^2_{\rho,{\rm eff}}\simeq \tfrac{1}{2}M^2+{\cal O}(M^2e^{-\sqrt{\frac{2}{3}}\varphi_{\rm inf}})~,
\end{equation}
where $\varphi_{\rm inf}$ is the nearly constant effective VEV of scalaron at the beginning of inflation, and $\rho$ is canonically normalized after taking into account the scalaron-dependent factor in its kinetic term. The Hubble scale $H$ of inflation is essentially given by the scalaron mass, $H^2 \sim M^2$, up to the $M^2e^{-\sqrt{\frac{2}{3}}\varphi_{\rm inf}}$-dependent terms that are suppressed during slow-roll inflation.

Once $\varphi$ falls onto the saddle point of the full scalar potential $V$ in the Einstein frame, corresponding to the local maximum of the $U$-potential, the $U$-term starts to dominate the energy density of the universe, with $\rho$ driving the second stage of inflation. However, if the initial velocity of $\rho$ is not large enough, the classical solution could stop at the saddle point. In this case, quantum effects (like quantum diffusion) are expected to destabilize the trajectory, so that the second inflation can proceed, and the trajectory can reach one of the Minkowski minima. In order to gain more control over the second inflationary stage and the power spectrum enhancement, we use the $c$-term because it can shift the saddle point  away from $\rho=0$ so that the inflationary trajectory avoids that point. Larger values of $c$ make the second stage of inflation shorter.

Next, we study the inflationary solutions numerically. It is convenient to write the scalar kinetic terms in the NLSM form as
\begin{equation}
    e^{-1}{\cal L}\supset -\tfrac{1}{2}G_{AB}\partial\chi^A\partial\chi^B~,
\end{equation}
where the scalars are denoted by $\chi^A=\{\varphi,\rho\}$, and the NLSM metric (in the field target space)  is $G_{AB}={\rm diag}\{1,e^{-\sqrt{2/3}\varphi}\}$ with $A,B=1,2$. The 2-dimensional NLSM target space is hyperbolic with a negative curvature.

Using the Friedmann-Lemaitre-Robertson-Walker (FLRW) metric of a flat universe, $g_{mn}={\rm diag}\{-1,a^2,a^2,a^2\}$, with the time-dependent cosmic scale factor $a(t)$, we get the scalar field equations of motion,
\begin{equation}
    D_t\dot\chi^A+3H\dot\chi^A+G^{AB}\partial_BV=0~,\label{KG_toy}
\end{equation}
where we have used $H=\dot a/a$. The $D_t$ is the covariant time derivative acting on a field space 
vector ${\cal V}^A$ as follows:
\begin{equation}
    D_t{\cal V}^A\equiv\dot\chi^B D_B{\cal V}^A=\dot{\cal V}^A+\dot\chi^B\Gamma^A_{BC}{\cal V}^C~.
\end{equation}
Here $\Gamma^A_{BC}$ are the field space (NLSM) Christoffel symbols,~\footnote{The non-vanishing Christoffel symbols in our model are $\Gamma^{\varphi}_{\rho\rho}=\fracmm{1}{\sqrt{6}}e^{-\sqrt{2/3}\varphi}$ and $\Gamma^{\rho}_{\varphi\rho}=\Gamma^{\rho}_{\rho\varphi}=-\fracmm{1}{\sqrt{6}}$.} and $D_B$ are the corresponding covariant derivatives. The $V$ in Eq.~\eqref{KG_toy} is the scalar potential in the Lagrangian \eqref{L_FN_toy} in the Einstein frame. More explicitly, the equations of motion \eqref{KG_toy} read
\begin{align}
    \ddot\varphi+\tfrac{1}{\sqrt{6}}e^{-\sqrt{\frac{2}{3}}\varphi}\dot\rho^2+3H\dot\varphi+\partial_\varphi V &=0~,\label{KG_toy_varphi}\\
    \ddot\rho-\sqrt{\tfrac{2}{3}}\dot\varphi\dot\rho+3H\dot\rho+e^{\sqrt{\frac{2}{3}}\varphi}\partial_\rho V &=0~.\label{KG_toy_rho}
\end{align}
The Friedmann-Einstein equations are
\begin{align}
    3H^2 &=\tfrac{1}{2}G_{AB}\dot\chi^A\dot\chi^B+V~,\label{Fried_1_toy}\\
    \dot H &=-\tfrac{1}{2}G_{AB}\dot\chi^A\dot\chi^B~.\label{Fried_2_toy}
\end{align}

We introduce the standard (Hubble flow) slow-roll parameters as
\begin{equation}
    \epsilon\equiv -\fracmm{\dot H}{H^2}~,\quad \eta\equiv\fracmm{\dot\epsilon}{H\epsilon}~,
\end{equation}
that are supposed to be small, $\epsilon\ll 1$, $|\eta|\ll 1$, during slow-roll inflation. The spectral index of scalar perturbations and the maximum value of the tensor-to-scalar ratio are given by
\begin{equation}
    n_s=1-2\epsilon-\eta~,~~~r_{\rm max}=16\epsilon~,
\end{equation}
being evaluated at the horizon exit of the pivot scale. Other useful quantities include the velocity $T^A$ and the acceleration (turn rate) $\omega^A$ unit vectors,
\begin{equation}
    T^A\equiv \dot\phi^A/\dot\phi~,~~~\omega^A\equiv D_tT^A~,
\end{equation}
where $\dot\phi\equiv\sqrt{G_{AB}\dot\phi^A\dot\phi^B}$. We denote the modulus of $\omega^A$ as $\omega\equiv\sqrt{G_{AB}\omega^A\omega^B}$.

As regards the parameters of our model, one of them can be considered free under the condition that the Starobinsky-like inflation at the CMB scales is not spoiled. For example, we can take $b$ as the free parameter with the restriction that its absolute value cannot be much larger than one. The other two parameters $c$ and $f$ can be used to control the duration of the second stage of inflation and the height of the power spectrum peak.

\begin{figure}
\centering
  \centering
  \includegraphics[width=.45\linewidth]{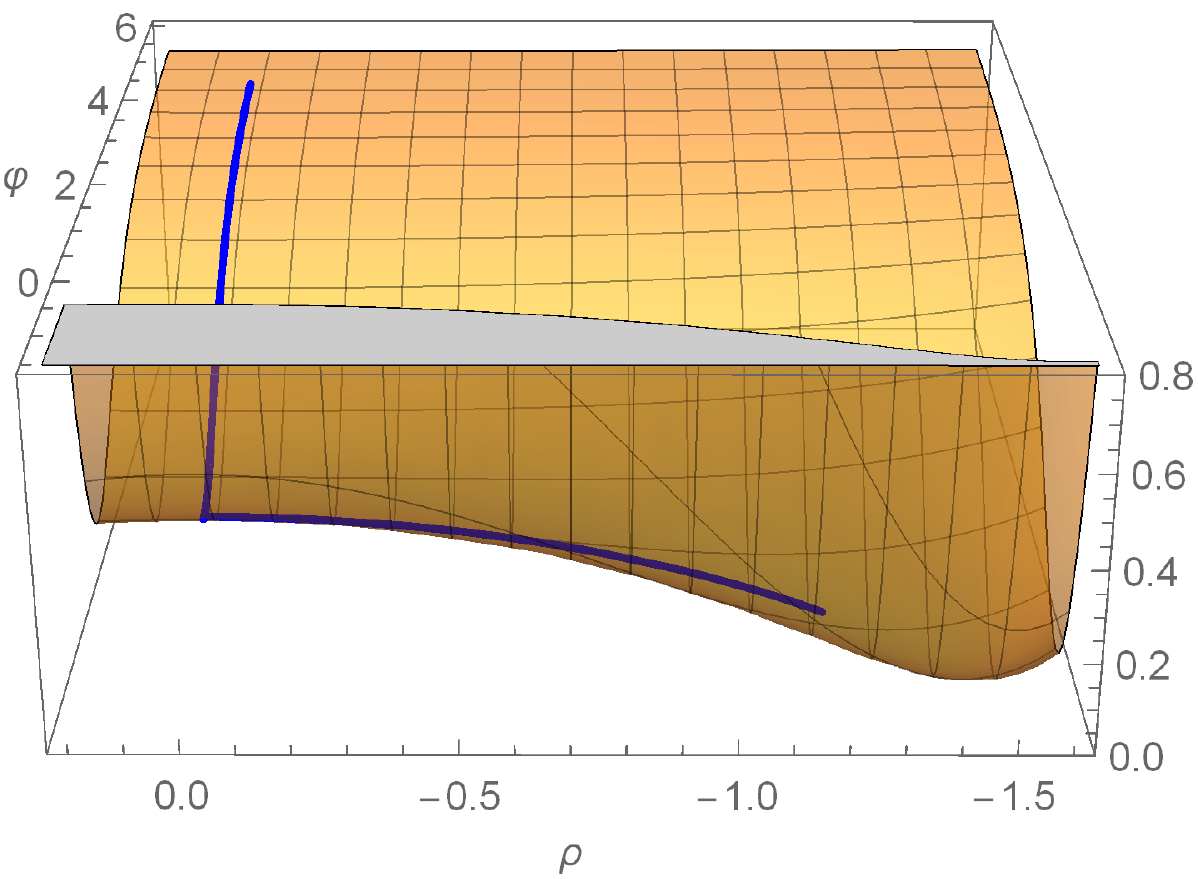}
\captionsetup{width=1\linewidth}
\caption{The inflationary trajectory superimposed on the scalar potential \eqref{L_FN_toy} with
 the parameters $b=-f=1$ and $c=10^{-4}$. The last $55$ e-folds are shown in blue.}\label{Fig_toy_traj}
\end{figure}

Let us take $b=-f=1$ in order to demonstrate the impact of the parameter $c$. For the second inflation to take place, $c$ must be small. For example, with $c=10^{-4}$ we obtain the solution to our equations \eqref{KG_toy_varphi}--\eqref{Fried_2_toy},~\footnote{In our numerical solutions we use the number of e-folds $N$ (defined by a solution to $\dot N=H$) as the dimensionless time variable.} whose trajectory is shown in Fig.~\ref{Fig_toy_traj} during the last $\Delta N$ e-folds of the observable inflation. In the Starobinsky inflation, the assumption of standard thermal history and the reheating temperature of about  $10^9$ GeV leads to $\Delta N\approx 55$ \cite{Bezrukov:2011gp}. We adopt this value throughout the paper.

We set the field initial conditions as 
\begin{equation}
    \varphi(N_0)=5.5~,~~~\rho(N_0)=0.1~,~~~\varphi'(N_0)=0.1~,~~~\rho'(N_0)=0.1~.\label{toy_ic}
\end{equation}
At the start of the numerical solution $\rho$ quickly settles at $\rho=0$ (due to its large effective mass) and stays there until the trajectory reaches the saddle point, as can be seen from Fig.~\ref{Fig_toy_traj}. After a few oscillations $\rho$ starts to move towards its negative VEV, which in this case is $\langle\rho\rangle\approx -\sqrt{2}$, while $\varphi$ is varying slowly. If $c$ were negative, $\rho$ would move towards $\langle\rho\rangle\approx +\sqrt{2}$ instead.

\begin{figure}
\centering
  \centering
  \includegraphics[width=.8\linewidth]{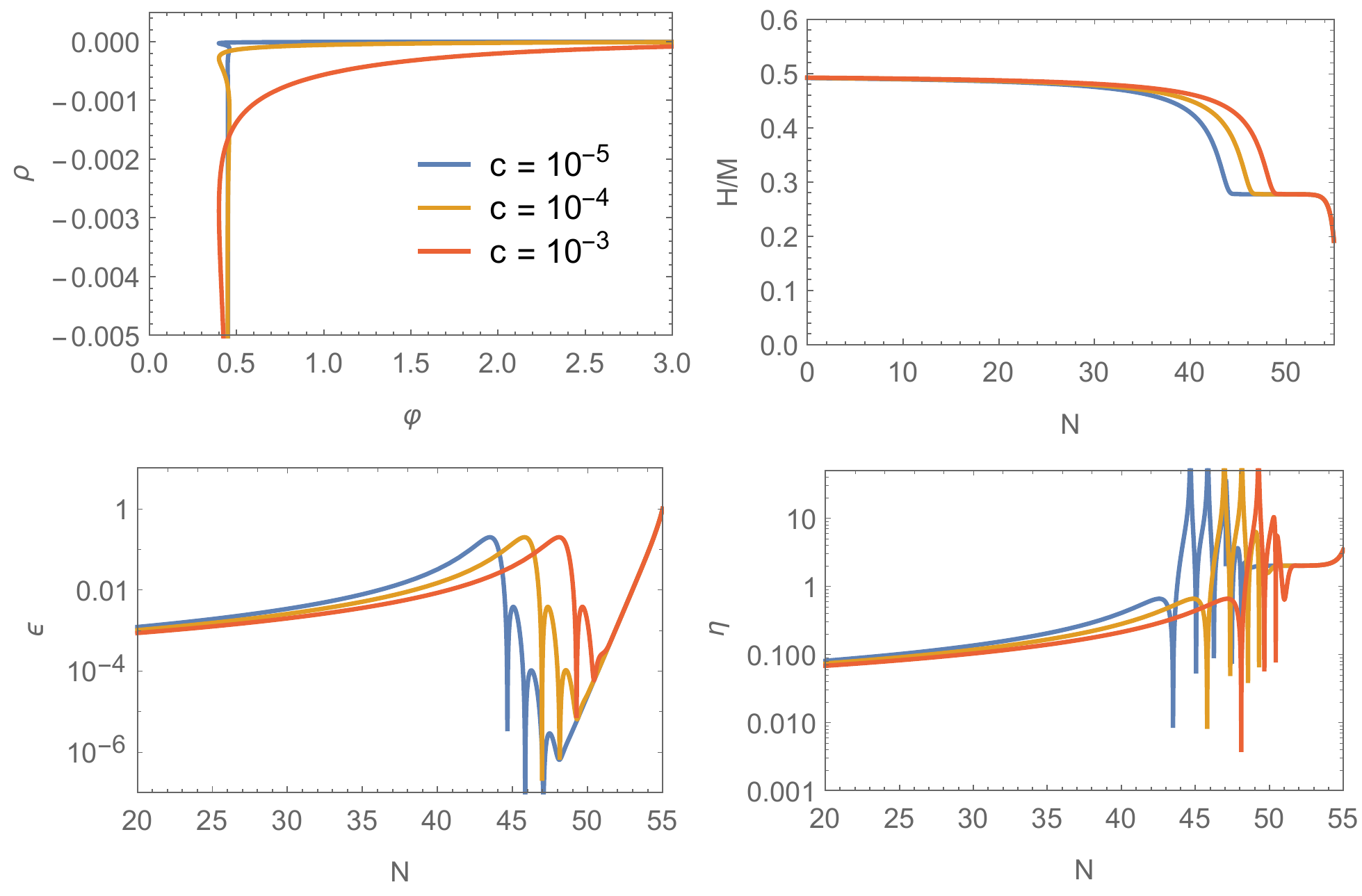}
\captionsetup{width=1\linewidth}
\caption{The solution to Eqs.~\eqref{KG_toy_varphi}--\eqref{Fried_2_toy} with the initial conditions \eqref{toy_ic} and $b=-f=1$. The two-field trajectory is on the upper left. The Hubble function during the last $55$ e-folds is on the upper right. The lower figures show the slow-roll parameters during the second ($\rho$-driven) stage of inflation. The colors correspond to $c=10^{-5}$ (blue), $c=10^{-4}$ (orange), and $c=10^{-3}$ (red).}\label{Fig_toy_SRH}
\end{figure}

In Figure \ref{Fig_toy_SRH} we show the inflationary trajectory on the $\varphi-\rho$ plane around the turning point (on the upper left), the Hubble function (on the upper right), and the slow-roll parameters around the second inflationary stage (on the lower figures) for the three values of the parameter $c=\{10^{-5},10^{-4},10^{-3}\}$ with $b=-f=1$. We define the start of the second stage of inflation as the moment when $\eta=1$ for the first time. We find that with the vanishing $c$ and the aforementioned initial conditions, the second stage  of inflation lasts over $60$ e-folds, with the power spectrum far exceeding unity at the peak. Therefore, we do not consider $c=0$.

\begin{figure}
\centering
  \centering
  \includegraphics[width=.85\linewidth]{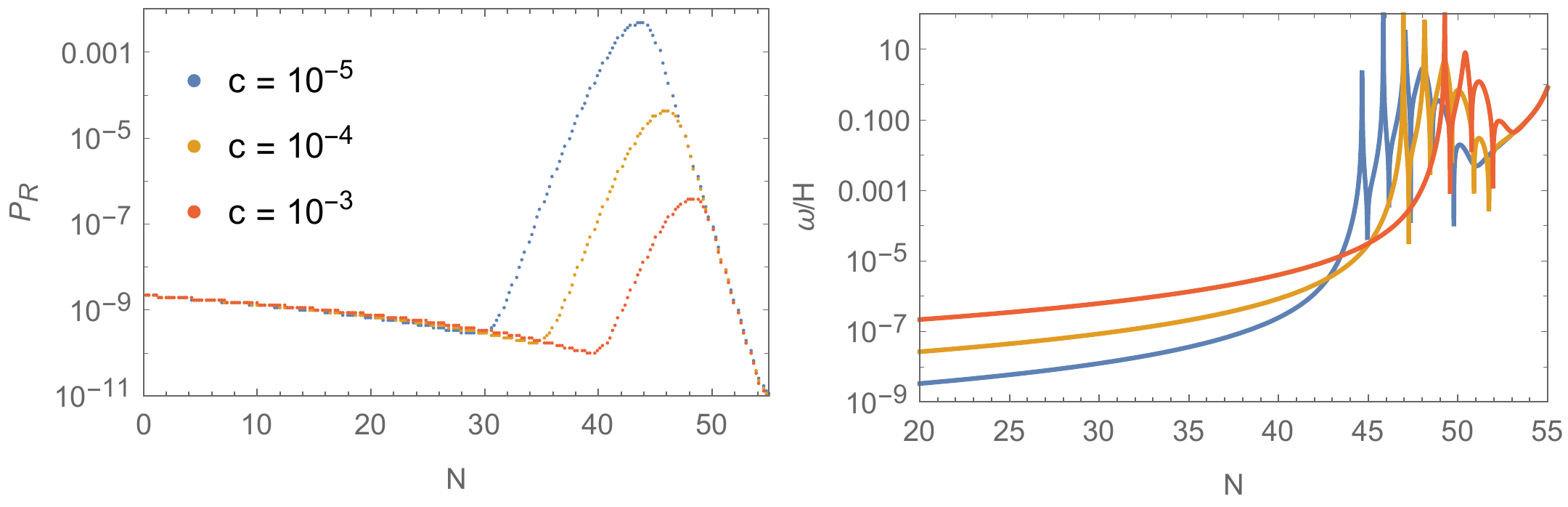}
\captionsetup{width=1\linewidth}
\caption{The power spectrum $P_R(k)$ against the e-folds number $N$ when the corresponding $k$-mode leaves the horizon ($k=a(N)H(N)$) for $b=-f=1$ and three reference values of $c$ (on the left). The turn-rate-to-Hubble ratio  $\omega/H$ is on the right. The color coding is the same as in Figure \ref{Fig_toy_SRH}.}\label{Fig_toy_Pkw}
\end{figure}

Table \ref{Tab_toy} shows the duration $\D N_2$ of the second stage of inflation, the inflationary observables, and the scalaron mass $M$ for three reference values of $c$. Figure \ref{Fig_toy_Pkw} shows the corresponding power spectrum $P_R(k)$ of scalar perturbations (on the left) and the turn rate modulus (on the right). For numerical computation of the power spectrum we use the transport method \cite{Mulryne:2010rp} and corresponding \textit{Mathematica} package described in \cite{Dias:2015rca}.

\begin{table}[hbt!]
\centering
\begin{tabular}{r r r r r}
\toprule
 $c$ & $\Delta N_2$ & $n_s$ & $r_{\rm max}$ & $M/M_P$ \\
\hline
$10^{-5}$ & $11.23$ & $0.9546$ & $0.0060$ & $1.60\times 10^{-5}$ \\
$10^{-4}$ & $8.93$ & $0.9569$ & $0.0054$ & $1.52\times 10^{-5}$ \\
$10^{-3}$ & $6.63$ & $0.9589$ & $0.0049$ & $1.45\times 10^{-5}$ \\\bottomrule
\hline
\end{tabular}
\captionsetup{width=1\linewidth}
\caption{The duration of the second inflationary stage $\Delta N_2$, inflationary observables $n_s$ and $r_{\rm max}$, and the scalaron mass $M$ for the three values of $c$ used in our examples.}
\label{Tab_toy}
\end{table}

Given the duration of the second stage of inflation and the slow-roll parameter $\epsilon$, we can  estimate the masses of PBH formed during the radiation domination (after restoring the reduced Planck mass $M_P$) as follows \cite{Pi:2017gih}:
\begin{equation}
    M_{\rm PBH}=\fracmm{M_P^2}{H_*}\exp\left(2\Delta N_2+\int_{N_*}^{N_{\rm exit}}\epsilon(N) dN\right)~,\label{toy_MPBH}
\end{equation}
where $N_{\rm exit}$ is the value of $N$ when the CMB scale $k=0.05~{\rm Mpc}^{-1}$ exits the horizon (in our notation we take $N_{\rm exit}=0$), $H_*$ and $N_*$ are the values at the end of the first inflation. In our  three examples with $c=\{10^{-5},10^{-4},10^{-3}\}$ and $b=-f=1$, Eq.~\eqref{toy_MPBH} shows that the PBH masses are $M_{\rm PBH}\sim\{10^9,10^7,10^5\}$ grams, respectively. They are too light because the PBH with masses smaller than $10^{16}$ g would have evaporated by now due to Hawking radiation. Larger PBH masses can be achieved by increasing the duration of the second stage of inflation $\Delta N_2$ because the formula \eqref{toy_MPBH} is most sensitive to this variable. The problem with this approach is a decrease of the value of $n_s$, see
e.g., Refs.~\cite{Aldabergenov:2020bpt,Aldabergenov:2020yok,Ishikawa:2021xya}. We can get the (model dependent) upper bound on $\Delta N_2$ from the $3\sigma$ CMB constraint on $n_s$ (the lower $3\sigma$ limit is around $0.945$ from Planck observations). 

\begin{figure}
\centering
  \centering
  \includegraphics[width=.9\linewidth]{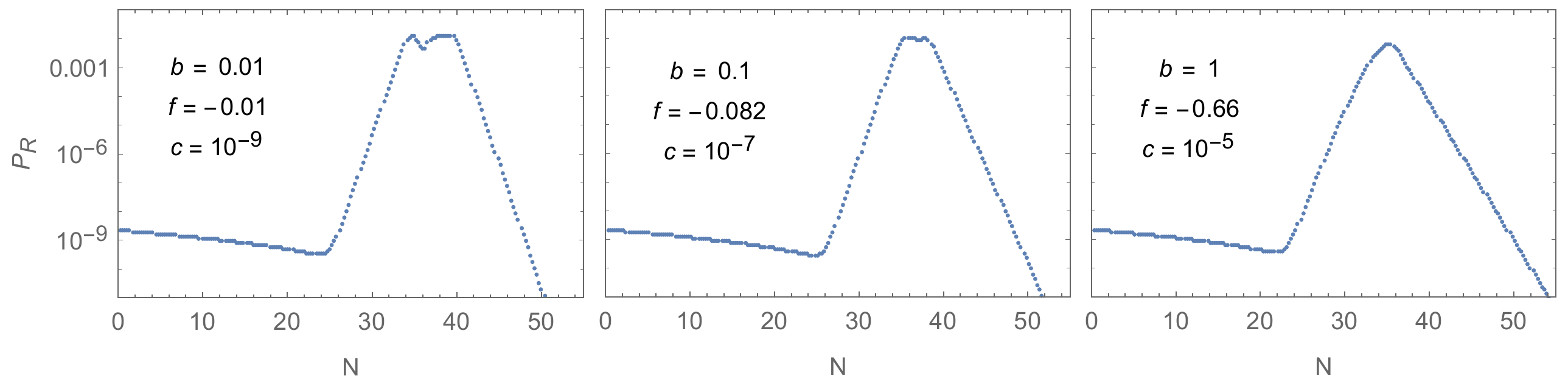}
\captionsetup{width=1\linewidth}
\caption{The power spectra corresponding to the sets of the parameters in Table \ref{Tab_toy_PBH}.}\label{Fig_toy_Pk_PBH}
\end{figure}

We find that in our simplified model $\Delta N_2=20$ leads to $M_{\rm PBH}\sim 10^{17}$ g. Hence, we target these values in our more advanced model in Sec.~4. Treating $b$ as an almost free parameter, we can check how the power spectrum behaves when $b$ changes. We choose the values $b=\{0.01,0.1,1\}$, and tune the parameters $f$ and $c$  in order to achieve $\Delta N_2\approx 20$ as well as the power spectrum peak of $P_R\sim 10^{-2}$, which are necessary for the PBH to account for DM in the Universe. A more precise value of PBH-DM, which also depends on the shape of the peak, can be found from numerical calculations of the PBH fraction. In our examples with three sets of the parameter values (leading to $\Delta N_2\approx 20$ and $M_{\rm PBH}\sim 10^{17}$ g), the predicted 
CMB observables (tilts) $n_s$ and $r_{\rm max}$ are given in Table \ref{Tab_toy_PBH}. The corresponding power spectra are shown in Fig.~\ref{Fig_toy_Pk_PBH}.

\begin{table}[hbt!]
\centering
\begin{tabular}{r r r r r}
\toprule
 $b$ & $f$ & $c$ & $n_s$ & $r_{\rm max}$ \\
\hline
$1$ & $-0.66$ & $10^{-5}$ & $0.9432$ & $0.0092$ \\
$0.1$ & $-0.082$ & $10^{-7}$ & $0.9463$ & $0.0081$ \\
$0.01$ & $-0.01$ & $10^{-9}$ & $0.9463$ & $0.0081$ \\\bottomrule
\hline
\end{tabular}
\captionsetup{width=1\linewidth}
\caption{Three sets of the parameter values used to produce the power spectrum peaks in Figure \ref{Fig_toy_Pk_PBH}. In all these cases we have $\Delta N_2\approx 20$ and $M_{\rm PBH}\sim 10^{17}$ g.}
\label{Tab_toy_PBH}
\end{table}

As can be seen from Fig.~\ref{Fig_toy_Pk_PBH}, smaller values of $b$ and $f$ lead to broader peaks featuring the oscillatory behavior. This may lead to a broad or a multipeak PBH mass distribution.

\section{Full model: three-field double inflation}\label{sec_full}

Having obtained insights into inflation and PBH production in our simplified model where the scalar $X$ was ignored and SUSY breaking did not occur, let us consider full dynamics and interactions of all the scalar fields involved, and search for SUSY breaking Minkowski vacua while keeping the two-stage inflation and the power spectrum enhancement of the simplified model as in the previous Section. For this purpose we add an extra term to the matter K\"ahler potential, while superpotential  remains the same:
\begin{align} \label{JOmega}
    J &=\phi\bar\phi-\tfrac{\lambda}{2}(\phi\bar\phi)^2~,\\
    \Omega &=b\phi+\tfrac{c}{2}\phi^2+\tfrac{f}{3}\phi^3~,
\end{align}
where we have four real parameters $\{\lambda,b,c,f\}$. The quartic term in the K\"ahler potential has been added for more flexibility. More specifically, we find that it can be used to control the power spectrum peak at the transition to the second inflationary stage (without the $\lambda$-term the peak can appear but is generally too small to accommodate a significant PBH abundance, see Sec.~6).

The master Lagrangian \eqref{L_FN_norm} in this case reads
\begin{equation}
	e^{-1}{\cal L}=\frac{1}{2}R-\frac{1}{2}\partial\varphi\partial\varphi-e^{-\sqrt{\frac{2}{3}}\varphi}\left(B\partial X\partial\overbar X+J_{,\phi\bar\phi}\partial\phi\partial\bar\phi\right)-V\label{L_m1}
\end{equation}
with the scalar potential in the Einstein frame as
\begin{equation}
    V=\fracmm{3M^2}{4B}\left(1-Ay\right)^2+y^2U~,
\end{equation}
where we have introduced the notation 
\begin{equation} \label{ydef}
y\equiv e^{-\sqrt{2/3}\varphi}~.
\end{equation}
The $A$, $B$, and $U$ are functions of $X,\overbar X$ and $\phi,\bar\phi$ with
\begin{equation}
    X=\tfrac{1}{\sqrt{2}}(\sigma+i\hat\sigma)~,~~~
    \phi=\tfrac{1}{\sqrt{2}}(\rho+i\hat\rho)~,
\end{equation}
where the hats are used to denote the imaginary parts (pseudo-scalars). To begin with, we set $\hat\sigma=\hat\rho=0$ for simplicity (we will show below that they are stabilized during and after inflation), and focus on the real parts of $X$ and $\phi$. We find from Eq.~\eqref{U_def_FN_choice} that
\begin{equation}
    A=1+\tfrac{1}{6}(\sigma^2-\rho^2)-\tfrac{11}{24}\zeta\sigma^4+\tfrac{1}{24}\lambda\rho^4~,\quad
    B=1-\zeta\sigma^2~,
\end{equation}
so that the scalar potential $U$ in the Jordan frame reads
\begin{align}
\begin{split}
    U/M^2\equiv\tilde U &=\fracmm{1}{12(1-\lambda\rho^2)}\left[2b+\sqrt{2}c\rho+f\rho^2+\sigma\rho\left(1-\frac{\lambda}{2}\rho^2\right)\right]^2\\
    &-\sigma\rho\left(b+\fracmm{c}{2\sqrt{2}}\rho+\fracmm{f}{6}\rho^2\right)+\frac{1}{2}\sigma^2\left(1-\frac{1}{6}\sigma^2-\frac{1}{6}\rho^2+\frac{3}{8}\zeta\sigma^4+\frac{\lambda}{24}\rho^4\right)~.\label{U_tilde}
\end{split}
\end{align}

Let us consider the early stage of inflation where $y\equiv e^{-\sqrt{2/3}\varphi}\ll 1$, and calculate the effective masses of the scalars $\sigma$ and $\rho$ at small $y$. Since $U$ is suppressed during this stage, and $A$ and $B$ are the functions of $X\overbar X$ and $\phi\bar\phi$, the effective masses of real and imaginary parts of $X$ and $\phi$ will be equal in the leading order with respect to $y$,~\footnote{The masses of $\hat\rho$ and $\hat\sigma$ in the SUSY-breaking Minkowski vacua are calculated in Table \ref{Tab_m1_F} of Sec.~\ref{sec_SUSY} for the specific parameter sets.}
\begin{align}
    m^2_{\sigma,{\rm eff}} &=m^2_{\hat\sigma,{\rm eff}}=\tfrac{3}{2}M^2(\zeta y^{-1}-2\zeta-\tfrac{1}{3})+{\cal O}(y)~,\\
    m^2_{\rho,{\rm eff}} &=m^2_{\hat\rho,{\rm eff}}=\tfrac{1}{2}M^2+{\cal O}(y)~.
\end{align}
Therefore, $\rho$ and $\hat\rho$ are stabilized during inflation, while the effective mass of $\sigma$ and $\hat\sigma$ depends not only on $\zeta$ but also on the value of $y$ (i.e. $\varphi$) during the first stage of inflation. For example, in the original Starobinsky model, the horizon exit of the CMB scale happens at approximately $\varphi=5.5$ or $y\approx 0.011$; in that case we find the condition $\zeta\gtrsim 0.004$. However, there is a stronger constraint coming from the requirement of a non-negative global minimum of the potential, $\zeta\geq 1/54\approx 0.019$ \cite{Aldabergenov:2020bpt}. In order to exclude possible meta-stable de Sitter minima, we choose a larger value, $\zeta=0.1$. In this case, we have at the horizon exit
\begin{equation}
    m^2_{\sigma,{\rm eff}}=m^2_{\hat\sigma,{\rm eff}}\simeq 12.9M^2+{\cal O}(y)~.
\end{equation}
Assuming that the other parameters of the model are of the order ${\cal O}(1)$ at most, the first stage of inflation is effectively a single-field inflation driven by the scalaron $\varphi$. Once $y$ becomes non-negligible, the Jordan frame potential $U$ will generically introduce a multifield dynamics involving $\sigma$ and $\rho$ that eventually settle at their minima, assuming $\langle\rho\rangle,\langle\sigma\rangle\neq 0$.

Next, we study the potential $U$ around the origin $\sigma=\rho=0$, where the second stage of inflation can take place. First we take $c=0$, and will later use $c\neq 0$  as a small perturbation in order to control the duration of the second stage of inflation. When ignoring $c$, the potential \eqref{U_tilde} becomes
\begin{align}
\begin{split}
    \tilde U &=\fracmm{1}{12(1-\lambda\rho^2)}\left[2b+f\rho^2+\sigma\rho\left(1-\frac{\lambda}{2}\rho^2\right)\right]^2\\
    &-\sigma\rho\left(b+\frac{f}{6}\rho^2\right)+\frac{1}{2}\sigma^2\left(1-\frac{1}{6}\sigma^2-\frac{1}{6}\rho^2+\frac{3}{8}\zeta\sigma^4+\frac{\lambda}{24}\rho^4\right)~.\label{U_tilde_c0}
\end{split}
\end{align}
At $\rho=\sigma=0$ the second derivatives of $\tilde U$ are
\begin{equation}
    \tilde U_{\rho\rho}=\frac{2}{3}b(f+b\lambda)~,~~~\tilde U_{\sigma\sigma}=1~,~~~\tilde U_{\rho\sigma}=-\frac{2}{3}b~,
\end{equation}
and the corresponding Hessian determinant is
\begin{equation}
    D(\tilde U)=\tilde U_{\rho\rho}\tilde U_{\sigma\sigma}-\tilde U_{\rho\sigma}^2=\frac{2}{3}b(f+b\lambda)-\frac{4}{9}b^2~.
\end{equation}
Since $\tilde U_{\sigma\sigma}>0$, there can be no local maximum. Therefore, in order  to support the second stage of inflation (with a graceful exit) we need a saddle point, i.e. one tachyonic direction around $\rho=\sigma=0$, which requires
\begin{equation}
    D(\tilde U)=\frac{2}{3}b(f+b\lambda)-\frac{4}{9}b^2<0~.\label{saddle_cond}
\end{equation}

We assign the following roles to each parameter of the model. First, we fix $\zeta=0.1$ in order to strongly stabilize the 
$\sigma$-direction both during and after inflation. The origin of the $\zeta$-dependent terms in the potential is from the second term of the function $N(X,\overbar X)$ in Eq. \eqref{FN_choice}. The parameter $b$ is treated as a free parameter, while $f$ can be fixed by the Minkowski vacuum equations once the other parameters are chosen. We will use $|c|\ll 1$ (or, when the other parameters are also small, we aim for the hierarchy $|c|\ll |b|,|f|,|\lambda|$) in order to control the duration of the second stage of inflation around $\sigma=\rho=0$, which will indirectly control the PBH masses. We find that $\lambda$ can be used to control the height of the power spectrum peak, i.e. the PBH abundance (and the corresponding GW density as well). Given these considerations, the condition for the saddle point \eqref{saddle_cond} is generically satisfied, so that we do not encounter any obstacles in that regard.

\begin{figure}
\centering
  \centering
  \includegraphics[width=.8\linewidth]{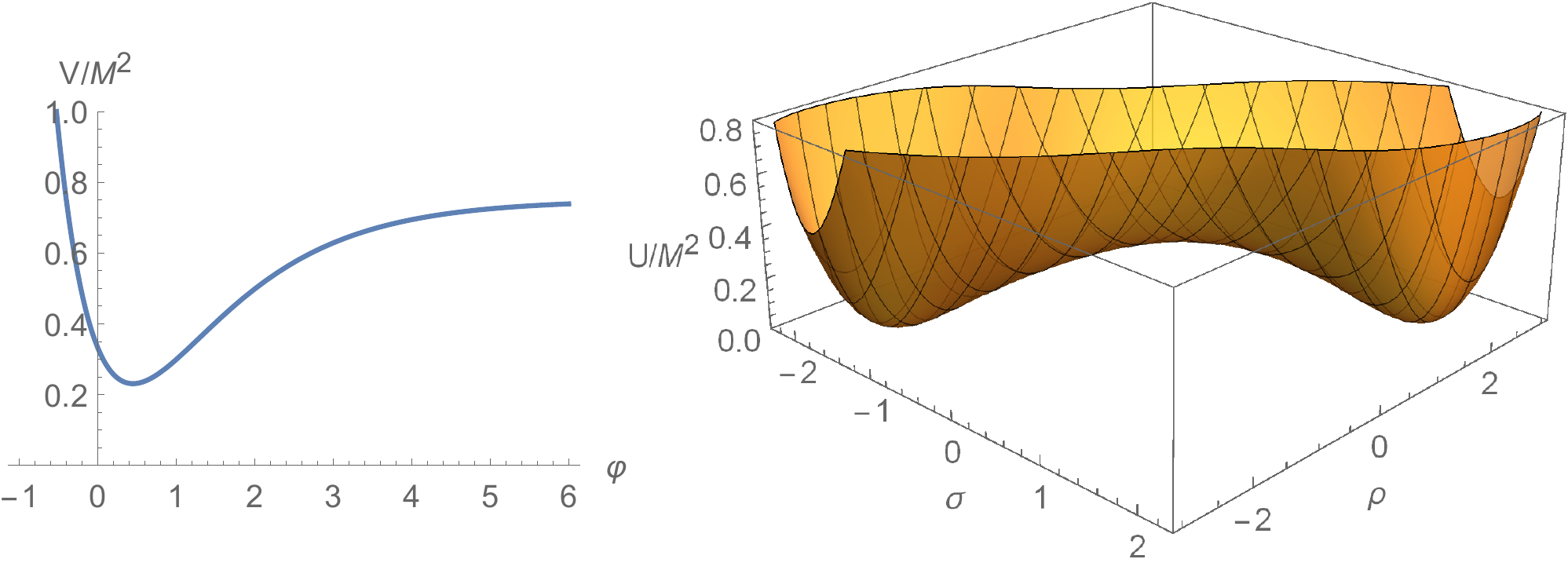}
\captionsetup{width=1\linewidth}
\caption{On the left: the Einstein frame scalar potential $V(\varphi)$ at $\rho=\sigma=0$. On the right: the Jordan frame potential $U(\rho,\sigma)$. The parameters are $b=1$, $f=0.404$, $c=0$, and $\lambda=0$.}\label{Fig_VU_eg}
\end{figure}

Let us numerically demonstrate the inflationary solutions and the behavior of the scalar power spectrum. First, we show the impact of the parameter $c$, while setting $b=1$ and $\lambda=0$. We plot the scalar potential for $c=0$ in Figure \ref{Fig_VU_eg}: the left plot shows $V(\varphi)$ at $\rho=\sigma=0$ (where $\varphi$ drives the first stage of inflation), while the right plot shows the Jordan frame potential $U(\rho,\sigma)$ (where $\rho$ and $\sigma$ drive the second stage of inflation). The potential $U$ has  the reflection symmetry $\{\rho,\sigma\}\rightarrow\{-\rho,-\sigma\}$ that is broken when $c\neq 0$. In other words, the saddle point of $U$ is shifted away from $\rho=\sigma=0$ in the presence of non-zero $c$. Therefore, by carefully tuning $c$, the length of the second stage of inflation can be controlled.

\begin{figure}
\centering
  \centering
  \includegraphics[width=.8\linewidth]{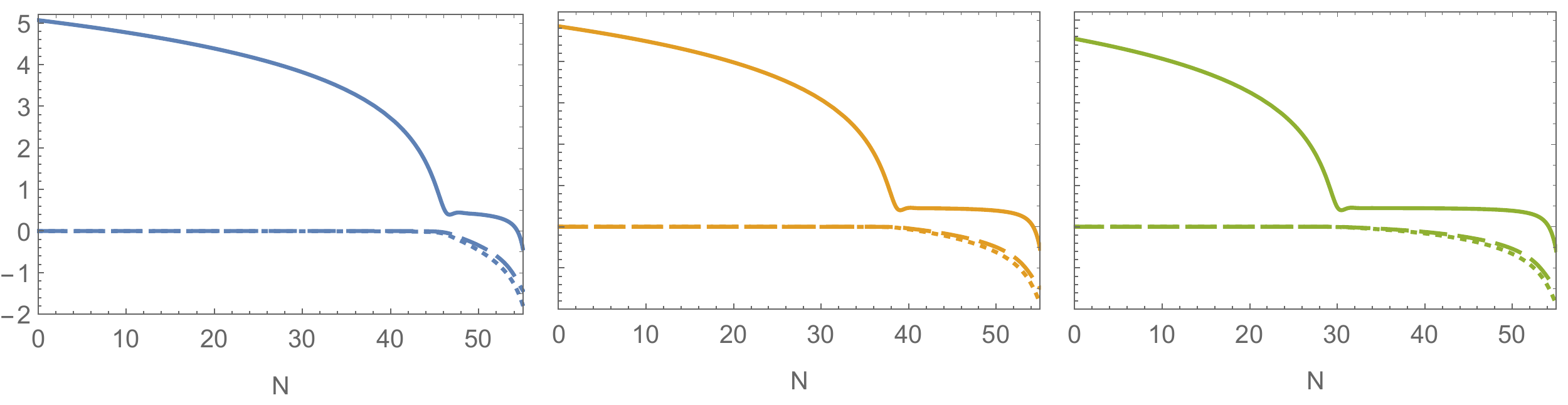}
\captionsetup{width=1\linewidth}
\caption{Evolution of $\varphi$ (solid line), $\rho$ (dotted) and $\sigma$ (dashed) during the last $55$ e-folds for the parameter sets \eqref{m1_cf_sets} with $c=0.1$ on the left, $c=0.03$ in the center, and  $c=0.01$ on the right.}\label{Fig_m1_c_sol}
\end{figure}

\begin{figure}
\centering
  \centering
  \includegraphics[width=.5\linewidth]{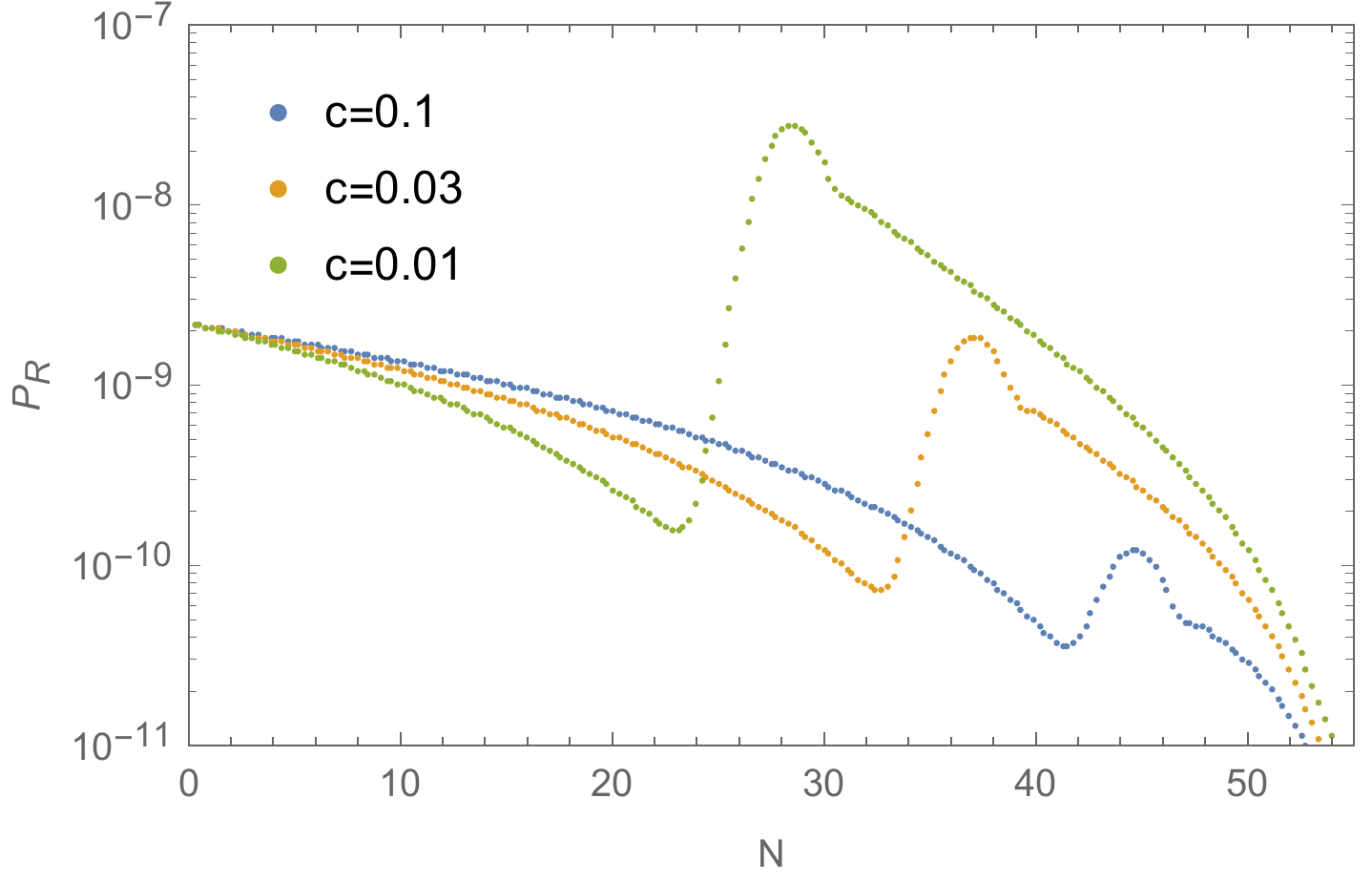}
\captionsetup{width=1\linewidth}
\caption{The power spectra $P_\car$ of scalar perturbations with the parameters \eqref{m1_cf_sets}. The horizontal axis shows the e-folds number when the corresponding mode $k$ crosses the horizon.}\label{Fig_m1_c_Pk}
\end{figure}

We get numerical solutions for different values of $c$. When varying $c$, we keep $b=1$ and $\lambda=0$ but adjust $f$ according to the vacuum equations. Choosing the sets of three parameters as
\begin{equation}
    \{c,f\}=\{0.1,0.4426\},\{0.03,0.4154\},\{0.01,0.4078\}~,\label{m1_cf_sets}
\end{equation}
we plot the Hubble function $H$, the turn rate $\omega/H$, and the slow-roll parameters $\epsilon,\eta$ during the last $55$ e-folds of inflation. Defining the end of inflation by $\epsilon=1$, and the end of the first stage by $\eta=1$ (generically, $\epsilon$ does not reach unity at that point), we can determine the duration $\Delta N_2$ of the second stage of inflation. The resulting $\Delta N_2$ together with the observables $n_s$, $r_{\rm max}$, and the mass parameter $M$ (to be found from the amplitude of scalar perturbations, $A_s\approx 2.1\times 10^{-9}$), for the parameter sets \eqref{m1_cf_sets} are shown in Table \ref{Tab_m1_c}. The evolution of the scalars $\varphi(N),\rho(N),\sigma(N)$ is shown in Figure \ref{Fig_m1_c_sol} for the same parameter sets. Our results for the power spectrum are shown in Figure \ref{Fig_m1_c_Pk}. As can be seen from Fig.~\ref{Fig_m1_c_Pk}, the peaks are too small to generate a sizable abundance of PBH.

\begin{figure}
\centering
  \centering
  \includegraphics[width=.7\linewidth]{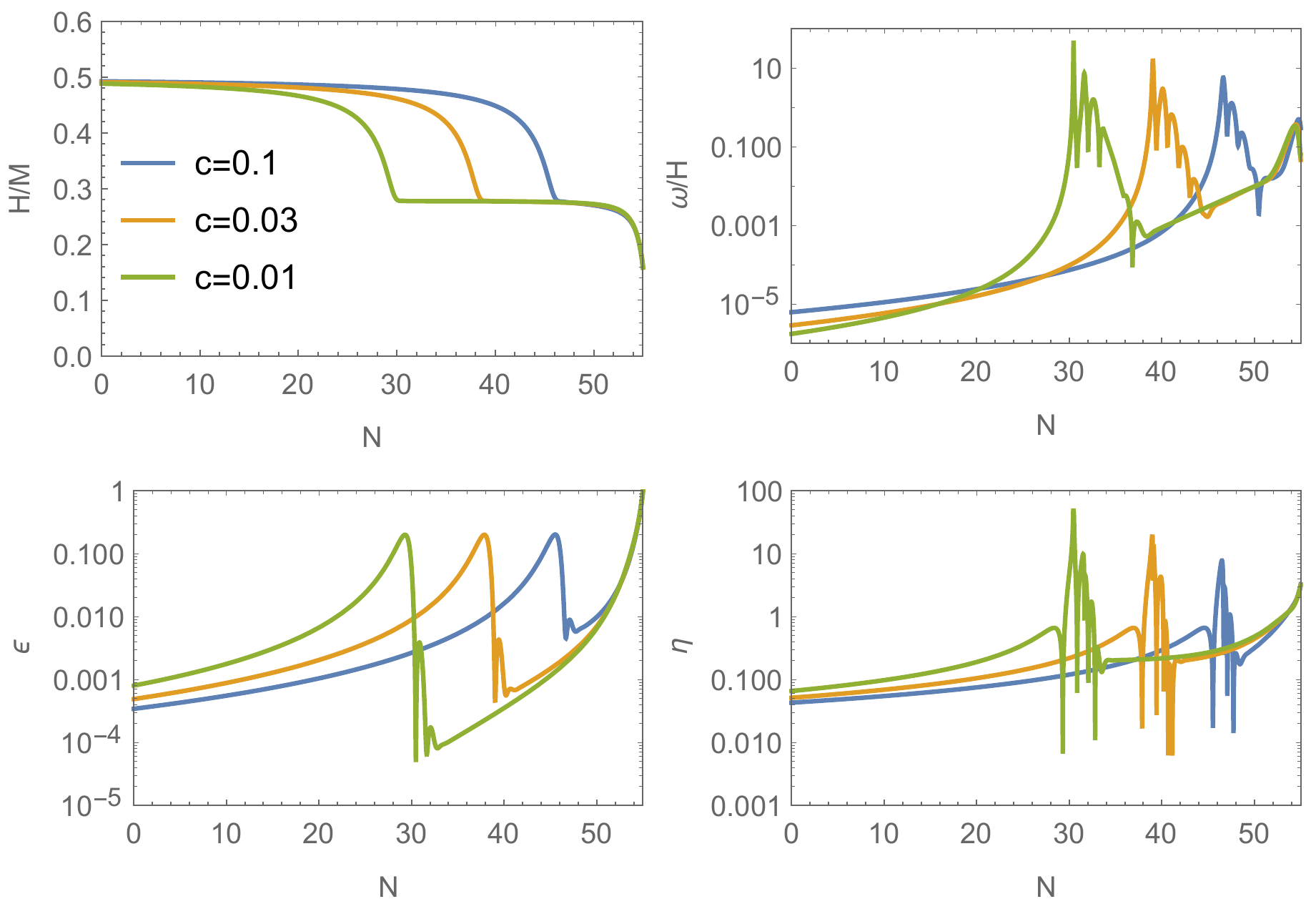}
\captionsetup{width=1\linewidth}
\caption{The Hubble function $H$, the turn rate $\omega/H$, and the slow-roll parameters $\epsilon,\eta$ during the last $55$ e-folds for the parameter sets \eqref{m1_cf_sets} in blue, orange, and green, respectively. The initial conditions for the fields are taken as $\varphi=5.5$, $\rho=\sigma=\varphi'=\rho'=\sigma'=0.1$ with the starting point around $20-30$ e-folds before the horizon exit which we define as $N=0$.}\label{Fig_m1_c_SR}
\end{figure}

\begin{table}[hbt!]
\centering
\begin{tabular}{r r r r r}
\toprule
 $c$ & $\Delta N_2$ & $n_s$ & $r_{\rm max}$ & $M$ \\
\hline
$0.1$ & $9.2$ & $0.9566$ & $0.0054$ & $1.53\times 10^{-5}$ \\
$0.03$ & $16.8$ & $0.9480$ & $0.0078$ & $1.83\times 10^{-5}$ \\
$0.01$ & $25.4$ & $0.9331$ & $0.0127$ & $2.36\times 10^{-5}$ \\\bottomrule
\hline
\end{tabular}
\captionsetup{width=1\linewidth}
\caption{The values of $\Delta N_2$, $n_s$, $r_{\rm max}$ and $M$ (in the Planck units) for the parameters \eqref{m1_cf_sets} with $b=1,\lambda=0$.}
\label{Tab_m1_c}
\end{table}

\begin{figure}
\centering
  \centering
  \includegraphics[width=1\linewidth]{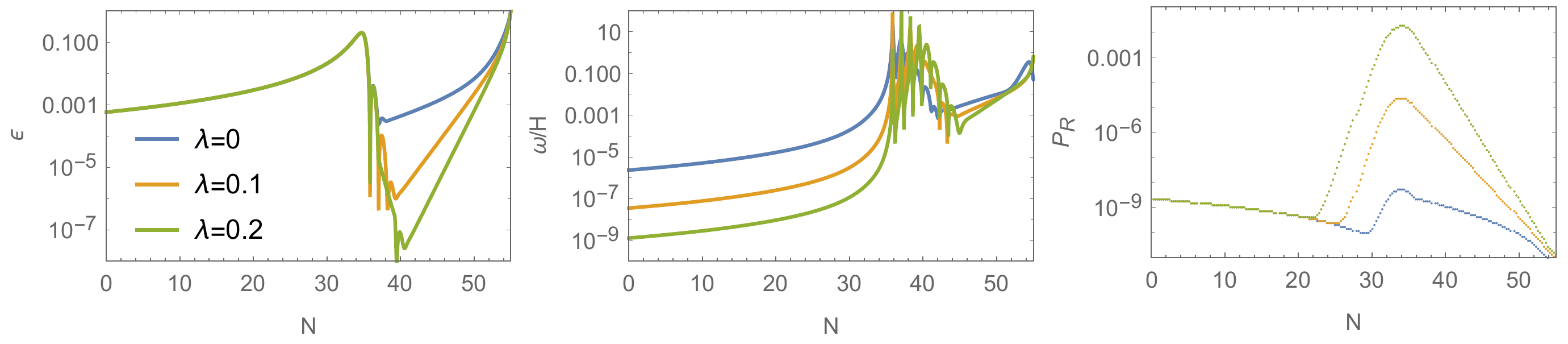}
\captionsetup{width=1\linewidth}
\caption{The slow-roll parameter $\epsilon$, the turn rate $\omega/H$ and the power spectrum $P_\car$ with the parameters in Table \ref{Tab_m1_lambda}. The color coding is shown on the left plot.}\label{Fig_m1_lambda_plots}
\end{figure}

After turning on the parameter $\lambda$, we show that it can enhance the peaks in the power spectrum leading to a significant amount of PBH. To demonstrate it, we set $b=1$, fix the length of the second stage of inflation as $\Delta N_2=20$ by adjusting $c$ and $f$, and vary $\lambda$. We take three different values of $\lambda=0$, $0.1$ and $0.2$, that imply the remaining parameters as shown in Table \ref{Tab_m1_lambda}. Since we have fixed $b=1$ and $\Delta N_2=20$, the inflationary observables $n_s$, $r$, and the mass scale $M$ are essentially the same for all the parameters chosen,
\begin{equation}
    n_s=0.9434~,~~~r_{\rm max}=~0.0092,~~~M=1.99\times 10^{-5}~.
\end{equation}
Although this value of $n_s$ is too small (outside the $3\sigma$ CMB constraints), it can be higher in some specific examples of PBH dark matter, as is shown in Sec.~\ref{sec_PBH}. The slow-roll parameter $\epsilon$, the turn rate $\omega/H$, and the power spectrum with the parameters in Table \ref{Tab_m1_lambda} are shown in Figure \ref{Fig_m1_lambda_plots}.

\begin{table}[hbt!]
\centering
\begin{tabular}{l | r r r}
\toprule
$\lambda$ & $0$ & $0.1$ & $0.2$ \\
$f$ & $0.4114$ & $0.0718$ & $-0.1968$ \\
$c$ & $0.0196$ & $2.9\times 10^{-4}$ & $1.05\times 10^{-5}$ \\
\bottomrule
\end{tabular}
\captionsetup{width=1\linewidth}
\caption{The three values of $\lambda$ and the corresponding parameters found by demanding $\Delta N_2=20$ after solving the vacuum equations with $b=1$.}
\label{Tab_m1_lambda}
\end{table}

\begin{figure}
\centering
  \centering
  \includegraphics[width=.85\linewidth]{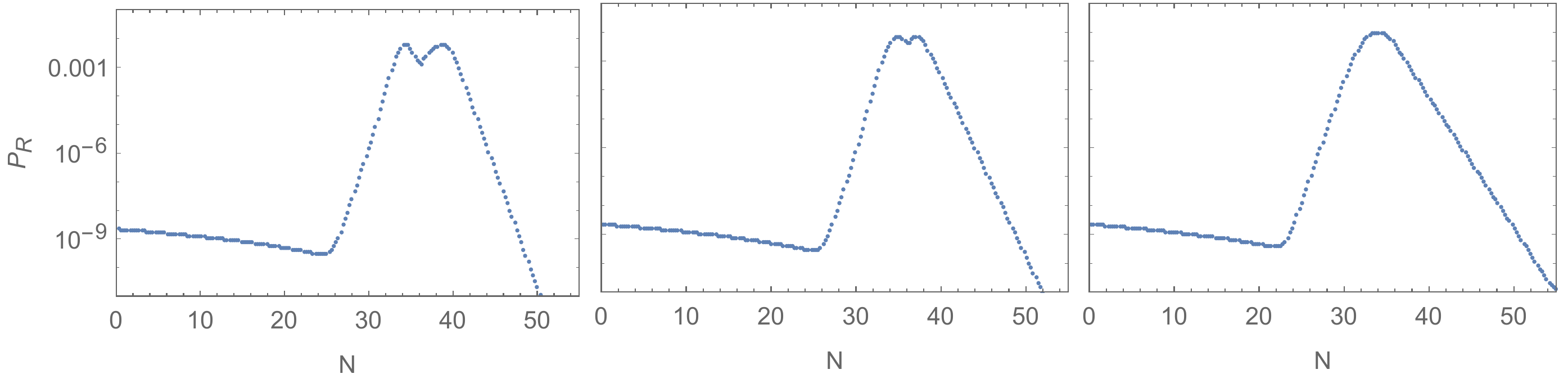}
\captionsetup{width=1\linewidth}
\caption{The power spectra $P_\car$ for three parameter sets in Table \ref{Tab_m1_b}: $b=0.01$ (on the left), $b=0.1$ (in the center), and $b=1$ (on the right).}\label{Fig_m1_b_Pk}
\end{figure}

Above we always set $b=1$ in order to demonstrate the impact of varying the parameters $c$ and $\lambda$, with $f$ being fixed by the Minkowski vacuum equations in each case. Finally, we study how the power spectrum peak behaves when varying $b$. We exclude large values of $b\gtrsim{\cal O}(10)$ because they spoil the first stage of inflation driven by $\varphi$, and consider $b=1,0.1,0.01$ as the examples. Having fixed again $\Delta N_2=20$ and having required the power spectrum to reach $P_\car\sim 10^{-2}$ (this is necessary to obtain the total PBH-to-DM fraction $\hat f_{\rm tot}=1$), we obtain the parameter values shown in Table \ref{Tab_m1_b}. In Figure \ref{Fig_m1_b_Pk}, the power spectra are displayed for those parameters. 

As can be seen from Fig.~\ref{Fig_m1_b_Pk}, our model can easily accommodate the power spectrum enhancement that can lead to PBH as the whole DM or as a significant part of it. It is worth noticing that the spectral tilt $n_s$ is slightly higher (for a better fit with CMB) for smaller values of $b$ in Table \ref{Tab_m1_b}.

\begin{table}[hbt!]
\centering
\begin{tabular}{r r r r r r r}
\toprule
 $b$ & $c$ & $f$ & $\lambda$ & $n_s$ & $r_{\rm max}$ & $M$ \\
\hline
$1$ & $1.4\times 10^{-5}$  & $-0.1717$ & $0.19$ & $0.9434$ & $0.0092$ & $2.00\times 10^{-5}$ \\
$0.1$ & $1.7\times 10^{-7}$  & $-0.03863$ & $0.27$ & $0.9463$ & $0.0082$ & $1.88\times 10^{-5}$ \\
$0.01$ & $1.8\times 10^{-9}$  & $-0.007098$ & $0.42$ & $0.9464$ & $0.0081$ & $1.87\times 10^{-5}$\\\bottomrule
\hline
\end{tabular}
\captionsetup{width=1\linewidth}
\caption{The parameters $c,f,\lambda$ for the three values of $b$ and $\Delta N_2=20$, and the corresponding inflationary predictions. The $M$ is in the Planck units.}
\label{Tab_m1_b}
\end{table}

As far as the PBH masses for $\Delta N_2=20$ are concerned, regardless of the parameters chosen we get  $M_{\rm PBH}\simeq 10^{17}$ g. The PBH mass function can be computed numerically, see Sec.~\ref{sec_PBH}.

\section{Spontaneous SUSY breaking in Minkowski vacuum}\label{sec_SUSY}

In the improved model of the preceding Section, the Minkowski vacua at $\sigma,\rho\neq 0$ spontaneously break SUSY. To demonstrate it, we transit to the dual picture in which the Einstein supergravity is coupled to three chiral matter superfields that we denote as $\Psi^I=\{{\bf T},{\bf S},\Phi\}$ where $\Phi$ is the same matter superfield introduced in Sec.~\ref{sec_setup}, containing the complex scalar $\phi=(\rho+i\hat\rho)/\sqrt{2}$. We will relate the remaining dynamical scalars $\varphi$ and $\sigma$ to the components of $\bf T$ and $\bf S$, and calculate the auxiliary F-fields of $\Psi^I$.

Our master superfield Lagrangian \eqref{L_master_1} can be dualized by introducing the chiral Lagrange multiplier superfield $\bf T$, and using the equivalent Lagrangian in the form
\begin{equation}
    {\cal L}=\int d^2\Theta 2{\cal E}\left\{-\tfrac{1}{8}(\overbar{\cal D}^2-8{\cal R})\big[N({\bf S},\overbar{\bf S})+J(\Phi,\overbar\Phi)\big]+{\cal F}({\bf S})+\Omega(\Phi)+6{\bf T}({\bf S}-{\cal R})\right\}+{\rm h.c.} \label{L_dual_1}
\end{equation}

After rescaling  
\begin{equation}
    {\bf S}\rightarrow M{\bf S}/\sqrt{12}~,~~~\Omega\rightarrow M\Omega/\sqrt{3}~,
\end{equation}
we can rewrite the Lagrangian \eqref{L_dual_1} to the standard form~\cite{Aldabergenov:2020bpt,Aldabergenov:2020yok} 
\begin{equation}
    {\cal L}=\int d^2\Theta 2{\cal E}\left[\tfrac{3}{8}(\overbar{\cal D}^2-8{\cal R})e^{-K/3}+W\right]+{\rm h.c.}~,\label{L_superfield_dual}
\end{equation}
where the K\"ahler potential $K$ and the superpotential $W$ are given by ({\it cf.} Refs~\cite{Ketov:2013dfa,Kallosh:2013xya}) 
\begin{align}
    K &=-3\log\left[{\bf T}+\overbar{\bf T}-\tfrac{1}{3}N({\bf S},\overbar{\bf S})-\tfrac{1}{3}J(\Phi,\overbar\Phi)\right]~,\label{Kael}\\
    W &=\sqrt{3}M{\bf S}\left({\bf T}-\tfrac{1}{2}\right)+\tfrac{1}{\sqrt{3}}M\Omega(\Phi)~. \label{supV}
\end{align}
We denote the leading field components of $\bf T$ and $\bf S$ as $T={\bf T}|$ and $S={\bf S}|$, where the complex scalar $S$ is in one-to-one correspondence with the scalar $X=\car|$ in the higher-derivative formulation \eqref{L_master_1} because varying Eq.~\eqref{L_dual_1} with respect to  $\bf T$ gives ${\bf S}=\car$.  The same rescaling of $\car$ was used in the preceding Sections. Therefore, we can parametrize $S$ in the same way as we did above for $X$,
\begin{equation}
    S=\tfrac{1}{\sqrt{2}}(\sigma+i\hat\sigma)~.
\end{equation}

Next, we identify scalaron $\varphi$ in the dual picture. It turns out to be
\begin{equation}
    e^{\sqrt{\frac{2}{3}}\varphi}=T+\overbar T-\tfrac{1}{3}N(S,\overbar S)-\tfrac{1}{3}J(\phi,\bar\phi)~.\label{T_par}
\end{equation}
Thus $\varphi$ comes from the real part of $T$ but also includes the functions $N$ and $J$. The imaginary part of $T$ is a pseudo-scalar corresponding to the effective scalar $D_mb^m$ in the higher-derivative picture (see Eq.~\eqref{L_master_comp}), it can always be stabilized and does not affect the inflationary dynamics \cite{Ketov:2013dfa}. We take the functions $J(\phi,\bar\phi)$ and $\Omega(\phi)$ 
as in Eq.~(\ref{JOmega}) together with
\begin{equation}
    N=S\overbar S-\tfrac{\zeta}{2}(S\overbar S)^2~,
\end{equation}
where we fix $\zeta=0.1$ in order to avoid possible instabilities/meta-stabilities during and after inflation.
In general, we do not have an upper bound on $\zeta$.

As regards the auxiliary $F$-fields, they can be found by using the standard formula
\begin{equation}
    F^I=-e^{K/2}K^{\bar JI}D_{\bar J}\overbar W~,\label{F-fields}
\end{equation}
where $K^{\bar JI}$ is the inverse of the K\"ahler metric given by $K_{I\bar J}=\partial_I\partial_{\bar J}K$, and the indices $I,J$ run over three chiral scalars $(T,S,\phi)$. The $D_I W$ is the K\"ahler-covariant derivative of the superpotential, $D_IW=\partial_IW+K_IW$. As is clear from Eq.~\eqref{F-fields} with the superpotential \eqref{supV}, if the $F$-fields are non-vanishing, they are always proportional to the  mass parameter $M$. Therefore, the SUSY breaking scale must be close to the inflation scale $\sim 10^{13}$ GeV in our model, unless the scalar VEVs become very small or/and the parameters of the matter superpotential $\Omega$ are very small (in the Planck units). 

Let us consider the concrete examples from the previous Section, namely, the parameter sets from Table \ref{Tab_m1_b}, which were shown to lead to a substantial enhancement in the scalar power spectrum. The corresponding numerical results for the $F$-fields and the gravitino mass $m_{3/2}$ in the Minkowski vacuum are shown in Table \ref{Tab_m1_F}, where we also include the masses of the pseudo-scalars $\hat\sigma$ and $\hat\rho$, demonstrating that they are not destabilized after inflation.

\begin{table}[hbt!]
\centering
\begin{tabular}{r r r r r r r}
\toprule
 $b$ & $\fracmm{|\langle F^T\rangle|}{MM_P}$ & $\fracmm{|\langle F^S\rangle|}{MM_P}$ & $\fracmm{|\langle F^\phi\rangle|}{MM_P}$ & $\fracmm{\langle m_{3/2}\rangle}{M}$ & $\fracmm{m_{\hat\sigma}}{M}$ & $\fracmm{m_{\hat\rho}}{M}$\\
\hline
$1$ & $0.11$  & $0.624$ & $1.631$ & $1.121$ & $0.25$ & $1.21$ \\
$0.1$ & $6\times 10^{-5}$  & $0.048$ & $0.155$ & $0.092$ & $0.77$ & $0.19$\\
$0.01$ & $3\times 10^{-8}$  & $0.048$ & $0.014$ & $0.007$ & $0.82$ & $0.02$ \\\bottomrule
\hline
\end{tabular}
\captionsetup{width=1\linewidth}
\caption{The SUSY breaking VEVs of the auxiliary $F$-fields, the gravitino mass $m_{3/2}$, and the masses of $\hat\sigma$ and $\hat\rho$ for different values of the parameter $b$. The values of other parameters can be found in Table \ref{Tab_m1_b}.}
\label{Tab_m1_F}
\end{table}

The super-heavy gravitino (as the LSP = the lightest superparticle) can also contribute to DM
\cite{Addazi:2017ulg,Addazi:2018pbg,Ketov:2019mfc}.

\section{PBH fraction of DM}\label{sec_PBH}

In this Section we calculate the PBH-to-DM density fraction by using the standard (Press-Schechter) formalism \cite{Press:1973iz}. The PBH masses, the production rate, and the density contrast are given by (as the functions of $k$) \cite{Inomata:2017okj,Inomata:2017vxo}
\begin{gather}
    M_{\rm PBH}(k)\simeq 10^{20}\left(\fracmm{7\times 10^{12}}{k~{\rm Mpc}}\right)^2{\rm g}~, \quad \beta_f(k)\simeq\fracmm{\sigma(k)}{\sqrt{2\pi}\delta_c}
    e^{-\fracmm{\delta^2_c}{2\sigma^2(k)}}~,\\
    \sigma^2(k)=\fracmm{16}{81}\int\fracmm{dq}{q}\left(\fracmm{q}{k}\right)^4e^{-q^2/k^2}P_\car(q)~,\label{PBH_productionE}
\end{gather}
respectively, where $\delta_c$ is a constant representing the density threshold for PBH formation, with the analytical estimate yielding $\delta_c\simeq 1/3$ \cite{Carr:1975qj}. Numerically, one finds a larger value, $0.41\lesssim\delta_c\lesssim 2/3$ \cite{Musco:2018rwt}, depending on the shape of the power spectrum. In terms of the above functions, the PBH fraction can be estimated as
\begin{eqnarray}
    \fracmm{\Omega_{\rm PBH}(k)}{\Omega_{\rm DM}}\equiv \hat f(k)\simeq\fracmm{1.2\times 10^{24}\beta_f(k)}{\sqrt{M_{\rm PBH}(k){\rm g}^{-1}}}~~.\label{f_PBH}
\end{eqnarray}

We assume the Minimal Supersymmetric Standard Model (MSSM) physical degrees of freedom resulting in the numerical factor $1.2$, though the difference with the Standard Model (where one finds $1.4$) is negligible for our purposes. The total PBH-to-DM fraction is given by the integral
\begin{eqnarray}
    \hat f_{\rm tot}=\int d(\log M_{\rm PBH})\hat f(M_{\rm PBH})~.
\end{eqnarray}

\begin{figure}
\centering
  \includegraphics[width=.6\linewidth]{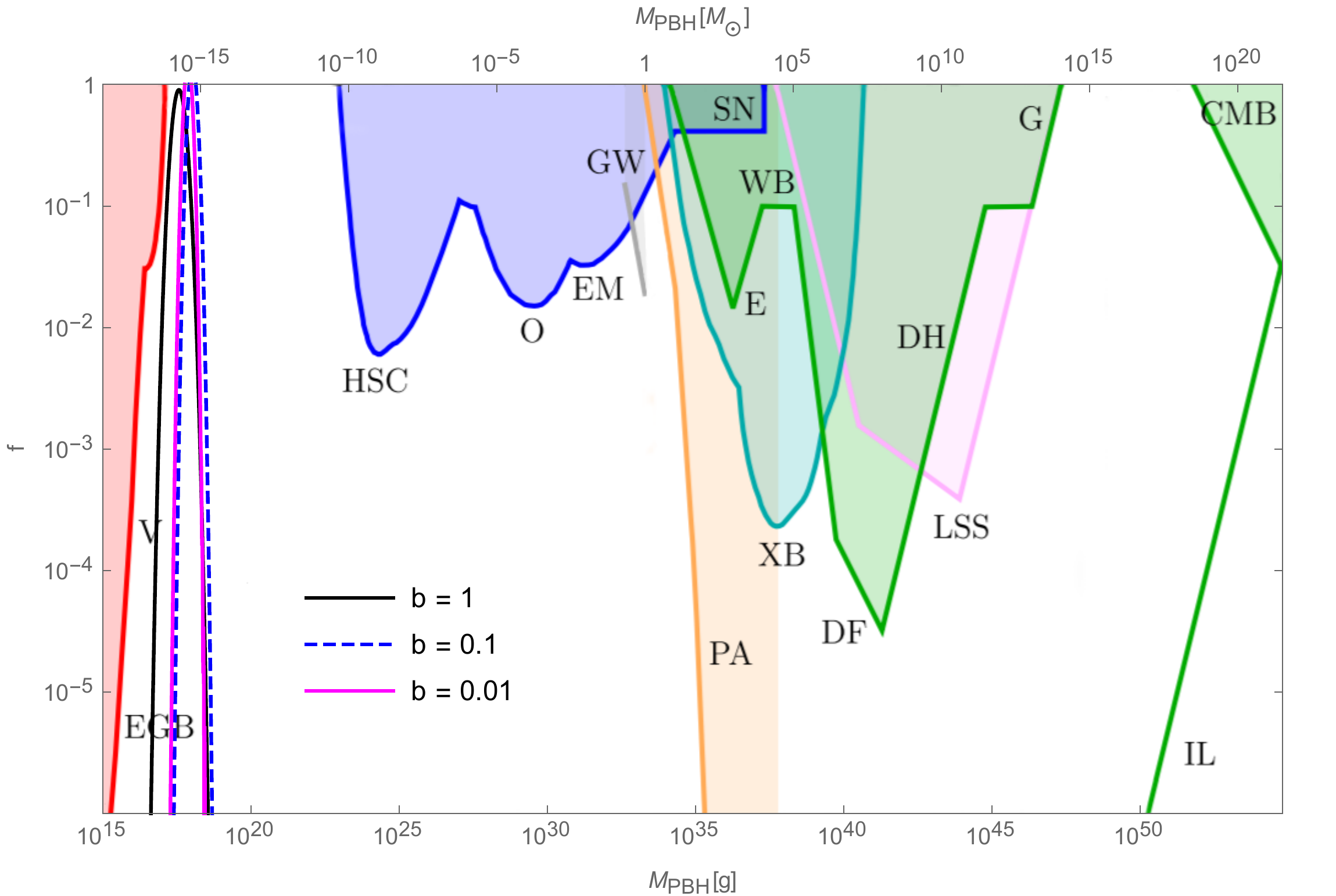}
\captionsetup{width=1\linewidth}
\caption{The PBH-to-DM fraction for the parameter sets A (solid black curve), B (dashed blue curve) and C (magenta curve) from Table \ref{Tab_f}, with $\hat f_{\rm tot}\approx 1$ in all cases. The background of observational constraints is taken from Refs.~\cite{Carr:2020gox,Carr:2020xqk}: from evaporation (red), lensing (blue), gravitational waves (gray), various dynamical effects (green), accretion (light blue), large-scale structure (pink), and CMB distortions (orange).}
\label{Fig_f}
\end{figure}

We normalize $k$ in accord with the scale $k=0.05~{\rm Mpc}^{-1}$ that leaves the horizon $55$ e-folds before the end of inflation. After choosing the reference value of the collapse threshold as $\delta_c=0.45$, we plot the resulting PBH fraction in Fig.~\ref{Fig_f} for the parameter sets A, B and C from Table \ref{Tab_f}. The parameter sets are tuned by the condition $\hat f_{\rm tot}\approx 1$. The corresponding values of $n_s$ are on the margin of the $3\sigma$ CMB constraint, and the PBH masses peak around $10^{18}$ g. The PBH observational constraints in Fig.~\ref{Fig_f} are taken from Refs.~\cite{Carr:2020gox,Carr:2020xqk} where they were obtained under the assumption of a monochromatic mass spectrum (we expect they are accurate enough for our purposes, since our peaks are narrow).

\begin{table}[hbt!]
\centering
\begin{tabular}{l r r r r r r r r}
\toprule
set & $b$ & $c$ & $f$ & $\lambda$ & $M$ & $n_s$ & $r_{\rm max}$ & $\Delta N_2$ \\
\hline
A & $1$ & $7.15\times 10^{-6}$  & $-0.327452$ & $0.254$ & $1.90\times 10^{-5}$ & $0.9462$ & $0.0083$ & $18.17$ \\
B & $0.1$ & $7.47\times 10^{-8}$  & $-0.047424$ & $0.31$ & $1.86\times 10^{-5}$ & $0.9467$ & $0.0080$ & $19.75$ \\
C & $0.01$ & $7.309\times 10^{-10}$  & $-0.009188$ & $0.52$ & $1.82\times 10^{-5}$ & $0.9478$ & $0.0077$ & $18.95$ \\\bottomrule
\hline
\end{tabular}
\captionsetup{width=1\linewidth}
\caption{The parameter sets (in Planck units) used to compute the PBH mass fractions in Fig.~\ref{Fig_f} and the inflationary observables.}
\label{Tab_f}
\end{table}
\vglue.2in

The parameter sets in Table \ref{Tab_f} may describe the whole dark matter by the PBHs of masses around $10^{18}$ g when accepting rather low values of the spectral tilt $n_s$, which are outside the $2\sigma$ CMB constraint but are still within the $3\sigma$ constraint. Smaller values of the parameter $b$ (as well as $c$ and $f$)  slightly increase $n_s$, however, decreasing the value of $b$ also decreases the mass of the pseudo-scalar $\hat\rho$,  see Table \ref{Tab_m1_F}.~\footnote{The values of the parameters $(c,f,\lambda)$ in Table \ref{Tab_f} differ from those in Tables \ref{Tab_m1_b} and \ref{Tab_m1_F} but they also lead to the SUSY breaking Minkowski vacua with the same (by the order of magnitude) estimates for the VEVs and masses as in Table \ref{Tab_m1_F} for given choices of $b$.} Hence, for a strong stabilization of all non-inflaton scalars during inflation larger values of $b$ are favored. As regards our choice of the parameters in Table \ref{Tab_f}, we use $b$ as a free parameter restricted to $b\lesssim 1$ for slow-roll inflation, the $f$ is fixed by the SUSY-breaking Minkowski vacuum equations, the $c$ helps to control the duration of the second inflation, and $\lambda$ provides the necessary power spectrum peak in order to describe the PBH dark matter (without the $\lambda$-term, the peak is too small).

\section{Induced gravitational waves}\label{sec_GW}

In this Section we calculate the scalar-induced gravitational wave (GW) density, and compare it to the expected sensitivity of the planned space-based GW interferometers.

The present-day GW density function $\Omega_{\rm GW}$ is given by \cite{Espinosa:2018eve,Bartolo:2018evs}
\begin{equation}
    \fracmm{\Omega_{\rm GW}(k)}{\Omega_{r}}=\fracmm{c_g}{72}\int^{\fracmm{1}{\sqrt{3}}}_{-\fracmm{1}{\sqrt{3}}}{\rm d}d\int^{\infty}_{\fracmm{1}{\sqrt{3}}}{\rm d}s\left[\fracmm{(s^2-\frac{1}{3})(d^2-\frac{1}{3})}{s^2+d^2}\right]^2 P_\zeta(kx)P_\zeta(ky)\left(I_c^2+I_s^2\right)~,
\end{equation}
where the constant $c_g\approx 0.3$ is taken in the case of the Minimal Supersymmetric Standard Model (MSSM). The radiation density $\Omega_{r}$ at present  is equal to $h^2\Omega_{r}\approx 2.47\times 10^{-5}$ \cite{Mather:1998gm}. Here $h$ is the reduced (present-day) Hubble parameter that we take as $h=0.67$ (ignoring the Hubble tension). The variables $x,y$ are related to the integration variables $s,d$ as
\begin{equation}
    x=\fracmm{\sqrt{3}}{2}(s+d)~,~~~y=\fracmm{\sqrt{3}}{2}(s-d)~,
\end{equation}
and the functions $I_c$ and $I_s$ are given by
\begin{align}
    I_c &=-4\int^{\infty}_0{\rm d}\eta\sin{\eta}\big\{ 2T(x\eta)T(x\eta)+\big[T(x\eta)+x\eta T'(x\eta)\big]\big[T(y\eta)+y\eta T'(y\eta)\big]\big\} ~,\\
    I_s &=4\int^{\infty}_0{\rm d}\eta\cos{\eta}\big\{ 2T(x\eta)T(x\eta)+\big[T(x\eta)+x\eta T'(x\eta)\big]\big[T(y\eta)+y\eta T'(y\eta)\big]\big\} ~,
\end{align}
where
\begin{equation}
    T(k\eta)=\fracmm{9}{(k\eta)^2}\left[\fracmm{\sqrt{3}}{k\eta}\sin\left(\fracmm{k\eta}{\sqrt{3}}\right)-\cos\left(\fracmm{k\eta}{\sqrt{3}}\right)\right]~,
\end{equation}
and $\eta$ is conformal time. The integration of $I_c$ and $I_s$ leads to \cite{Espinosa:2018eve}, 
\begin{align}
    I_c&=-36\pi\fracmm{(s^2+d^2-2)^2}{(s^2-d^2)^3}\theta(s-1)~,\\
    I_s&=-36\fracmm{s^2+d^2-2}{(s^2-d^2)^2}\left[\fracmm{s^2+d^2-2}{s^2-d^2}\log\left|\fracmm{d^2-1}{s^2-1}\right|+2\right]~,
\end{align}
where $\theta$ is the Heaviside step function.

\begin{figure}
\centering
  \includegraphics[width=.6\linewidth]{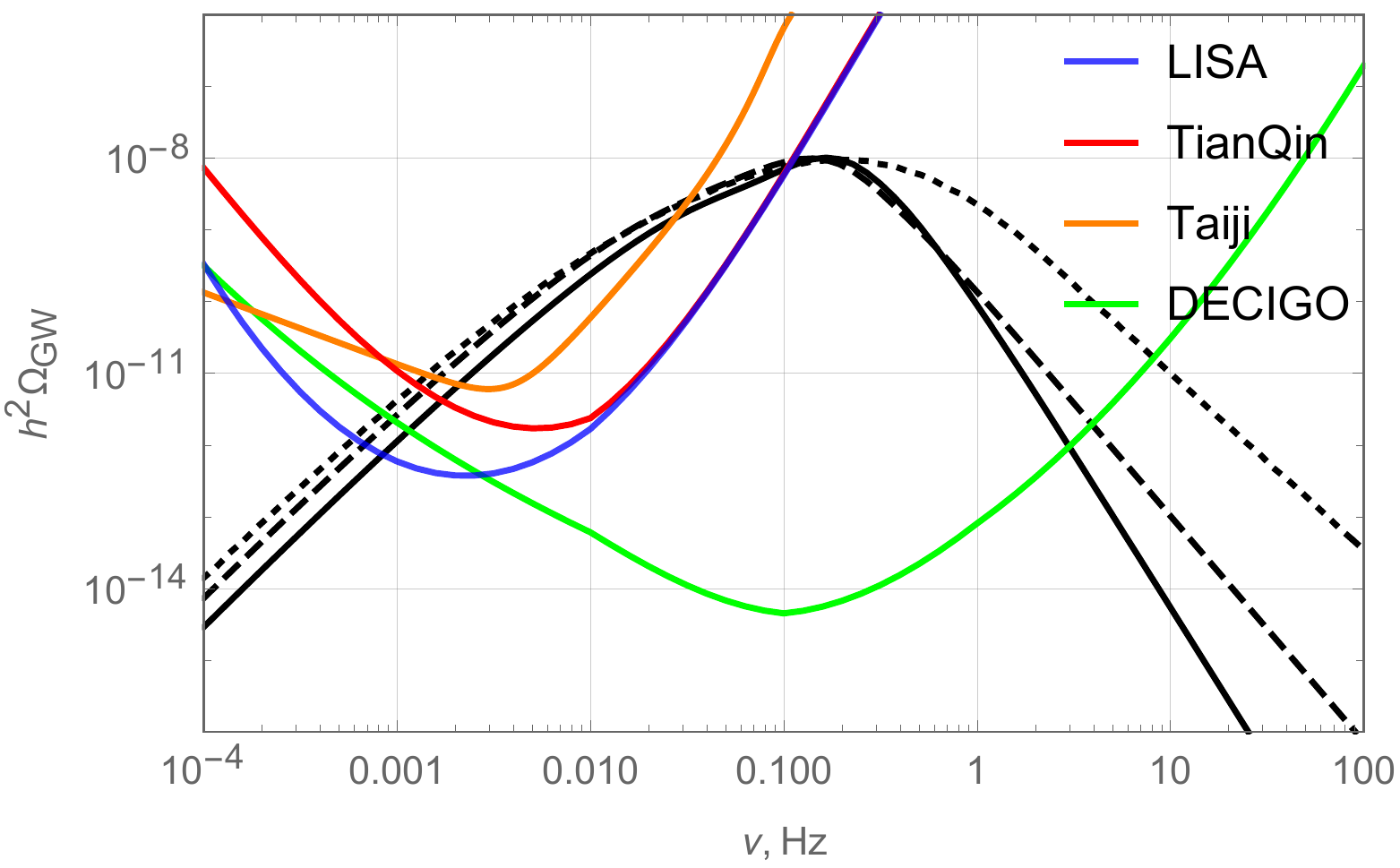}
\captionsetup{width=1\linewidth}
\caption{The GW density for the parameter sets in Table \ref{Tab_f}: set A (dotted curve), set B (dashed curve) and set C (solid curve).}
\label{Fig_GW}
\end{figure}

The GW density can be numerically computed for a given power spectrum. By using the power spectra with the parameter sets in Table \ref{Tab_f} we plot the density $\Omega_{\rm GW}(\nu)$ in terms of frequency 
$\nu=k/(2\pi)$ in Fig.~\ref{Fig_GW} together with the expected sensitivity curves~\footnote{The parameters and the noise models for LISA \cite{LISA}, TianQin \cite{TQ}, Taiji \cite{TAIJI,Ruan:2018tsw}, and DECIGO \cite{DEC} are used to get the sensitivity curves, respectively.} for several space-based GW experiments. As can be seen from Fig.~\ref{Fig_GW}, should a significant fraction of dark matter be formed by PBH, the corresponding induced gravitational wave background can be tested by the future space-based GW interferometers.

\section{Conclusion}

In this paper we proposed and studied the modified (Starobinsky-like) old-minimal-type supergravity coupled to a chiral matter superfield, that can {\it simultaneously} describe multifield inflation, PBH formation and DM, and spontaneous SUSY breaking after inflation in a Minkowski vacuum. 
First, we investigated a simplified construction with two dynamical scalars, in order to reveal the qualitative features of the double inflation scenario in our setup, where we ignored the complex scalar $X$ of the supergravity multiplet. We demonstrated that the first stage of inflation is driven by the Starobinsky scalaron, whereas the second stage (after a saddle point) is driven by the (real part of) matter scalar $\phi$
with the canonical K\"ahler potential and a Wess-Zumino superpotential including a linear term. At the onset of the second inflation, the scalar power spectrum is significantly enhanced, thus leading to the PBH formation once the curvature perturbations reenter the horizon during the radiation era. However, SUSY is restored after inflation in the simplified model.

Our full supergravity model was studied in Sec.~\ref{sec_full} where we derived the inflationary dynamics of all physical scalars. We found that the scalars have different roles: the scalaron $\varphi$ is a driver of the  first stage of inflation where the CMB scale exits the horizon, whereas the second stage of inflation is driven by a combination of $\sigma$ (the real part of $X$) and $\rho$ (the real part of the chiral matter scalar $\phi$). Like the simplified model, beginning of the second inflationary stage gives rise to an enhancement (peak) of the power spectrum. However, in order to achieve the required enhancement of $\co(10^7)$ for a substantial PBH production, we need to employ an additional term $\sim |\phi|^4$ in the K\"ahler potential. 

In Section \ref{sec_SUSY} we showed that in our models SUSY is generically spontaneously broken in the vacuum after inflation, while one of the parameters can be fixed to achieve the vanishing cosmological constant. We also numerically computed the masses of the imaginary scalars $\hat\sigma$ and $\hat\rho$ (from $X$ and $\phi$, respectively) around the vacuum, and found them close to the inflationary Hubble scale as long as the superpotential parameters are not too small. During inflation, the scalars $\hat\sigma$ and $\hat\rho$ are stable and have larger effective masses (provided the superpotential parameters are not too small). There is also an additional real scalar -- the scalar superpartner of scalaron -- that also has a similarly large (non-tachyonic) effective mass during and after inflation. In the initial higher-curvature formulation of our supergravity models, the extra scalar is related to the divergence $D_m b^m$ of the vector field $b_m$ belonging to the old-minimal supergravity multiplet, see Ref.~\cite{Ketov:2013dfa} for more. To summarize, we have six real scalars in our models, with the three of them being stabilized and the other three participating in inflation.

In Section \ref{sec_PBH} we gave some specific examples where the PBHs may describe the whole DM at present. The current observational constraints on PBH allow only the limited mass range for the whole PBH-DM, see Fig.~\ref{Fig_f}, between $10^{17}$ g and $10^{23}$ g. Having insisted on the PBH masses beyond the Hawking evaporation limit 
of $10^{15}$ g, needed for the PBH-DM in the current Universe, we got the rather low CMB tilts $n_s$ in our model,
outside the $2\sigma$ but within the $3\sigma$, as e.g., in Table 7.  This shortcoming motivates further generalizations of our model, e.g., by using more flexible $\alpha$-attractors \cite{Galante:2014ifa} for inflation and more general matter couplings in the supergravity framework (work in progress). It is also worth mentioning that our model avoids overproduction of PBHs in the given region of the parameter space ({\it cf.} Ref.~\cite{Bassett:2000ha}), while our results agree with other independent studies along similar lines \cite{Geller:2022nkr}.

SUSY breaking after inflation is known to be difficult in the modified supergravity, see e.g., Refs.~\cite{Addazi:2017ulg,Aldabergenov:2020bpt}. We found that the simplest Polonyi mechanism of spontaneous SUSY breaking does not work in the minimal setup for our supergravity models with the canonical K\"ahler potential and linear superpotential of matter. However, in Sec.~\ref{sec_SUSY}, it was shown to be possible  after a modification of the matter superpotential by Wess--Zumino terms and/or K\"ahler potential by a quartic term (both are needed for PBH-DM scenario).  The scale of SUSY breaking in our supergravity models is generally very high, under the GUT and inflation scales, which is reflected in the gravitino mass shown in Table \ref{Tab_m1_F}, that is close to the value of the mass parameter $M$.

Our supergravity model in this paper apparently favors the {\it composite} nature of DM as a mixture of PBH
and heavy gravitino as the lightest SUSY particle (LSP) because  the massive gravitino LSP is also a natural candidate for DM in supergravity theories with spontaneous SUSY breaking \cite{Addazi:2017ulg}. The composite DM significantly relaxes fine-tuning needed for the whole PBH-DM; see also Refs.~\cite{Dudas:2017rpa,Dudas:2017kfz} for a possible consistency of this scenario with the MSSM.

In Section \ref{sec_GW} we derived the second-order gravitational wave background induced by the enhanced scalar perturbations, and confirmed that those GW may be accessible by the future space-based GW experiments, as is expected from the (low-mass) PBH-DM scenarios \cite{Bartolo:2018evs,Inomata:2021uqj,Vaskonen:2020lbd}. The GW detected by the ground-based gravitational interferometers (LISA/Virgo/KAGRA) from the black hole mergers do not contain information about the origin and evolution of the black holes because such information is washed up during the mergers. For instance, in the case of LISA, the induced GW frequency should not be far away from 3.4 mHz, which implies the PBH masses of the order $10^{-12}M_{\odot}\sim  10^{21}$ g \cite{Bartolo:2018evs} that is close to the PBH masses we got from our modified supergravity models.

The main takeaway from our study reads: inflation, PBH formation, PBH-DM, and SUSY breaking can be unified in the supergravity framework and directly affect each other. In our model, the Polonyi-like SUSY-breaking field $\phi$ participates in the second stage of inflation that, in turn, generates large fluctuations and an enhancement in the scalar power spectrum, leading to the PBH formation. Demanding a strong stabilization of the non-inflaton scalars in our model leads to larger values of the parameter $b$ (close to one) that, in turn, constrains the SUSY breaking parameters $\langle F\rangle/M_P$ and $\langle m_{3/2}\rangle$ within one order of magnitude from the inflationary Hubble scale controlled by the mass parameter $M$ of the $R^2$ supergravity. The high SUSY breaking scale also helps us to avoid the gravitino and Polonyi/moduli problems \cite{Addazi:2017ulg}.

The $3\sigma$ constraints on the CMB tilt $n_s$ in our model imply that the possible PBH masses are limited to the lower values in the whole-PBH-DM window, i.e. to $10^{17}\div 10^{18}$ g,  as we found. Larger PBH masses further decrease $n_s$, while smaller PBH masses are excluded by the Hawking evaporation constraint. 


\section*{Acknowledgements}

YA was supported by the Chulalongkorn University CUniverse research promotion project (grant CUAASC), and the Thailand science research and innovation fund under the project CU$\_$FRB65$\_$ind (2)$\_$107$\_$23$\_$37. 

AA was supported by the Talent Scientific Research Program of College of Physics, Sichuan University, under the grant No.~1082204112427, the Fostering Program in Disciplines Possessing Novel Features for Natural Science of Sichuan University,  under the grant No.~2020SCUNL209, and the 1000 Talent Program of Sichuan province 2021.

SVK was supported by Tokyo Metropolitan University, the Japanese Society for Promotion of Science under the grant No.~22K03624, the World Premier International Research Center Initiative (MEXT, Japan), and the Tomsk Polytechnic University development program Priority-2030-NIP/EB-004-0000-2022. 

SVK acknowledges correspondence and discussions with G.~Dvali, C. Germani, D.I.~Kaiser, F.~Kuhnel, O.~Lechtenfeld, K.-I. Maeda, S.~Pi, M.~Sasaki and S.~Tsujikawa.

\providecommand{\href}[2]{#2}\begingroup\raggedright\endgroup

\end{document}